\documentclass{article}

\usepackage{amsmath}
\usepackage{amsfonts}
\usepackage{amssymb}
\usepackage{bm,color}
\usepackage{graphicx}
\usepackage{indentfirst}
\usepackage{array}
\usepackage{subfigure}
\usepackage{algorithm}
\usepackage{algorithmic}
\usepackage{enumitem}
\usepackage{authblk}
\usepackage{fullpage}

 \newcommand{\h}[1]{\mathbf{#1}}

\newcommand{\tensor}[1]{\mathbf{#1}}
\newcommand{\trans}{^\mathsf{T}}
\newcommand{\dif}{\mathrm{d}}

\newcommand{\ba}{\bm\alpha}
\newcommand{\sign}{\mbox{sign}}
\newcommand{\bx}{\bm\xi}
 
\newcommand{\ve}{\varepsilon}
\newcommand{\mn}{\mathbb{N}}
\newcommand{\mr}{\mathbb{R}}
\newcommand{\tc}{\tilde{\bm c}}

\DeclareMathOperator*{\argmin}{arg\,min}

\title{
A general framework of rotational
sparse approximation in uncertainty quantification}
\date{}

\author[1]{Mengqi Hu}
\author[2]{Yifei Lou}
\author[3]{Xiu Yang\thanks{Corresponding author: xiy518@lehigh.edu}}
\affil[1,2]{Department of Mathematical Sciences, University of Texas  Dallas,
Richardson, TX 75080, USA}
\affil[3]{Department of Industrial and Systems Engineering, Lehigh University,
PA 18015, USA}

\begin{document}
\maketitle
\begin{abstract}
This paper proposes a general framework to estimate coefficients of generalized
  polynomial chaos (gPC) used in uncertainty quantification via rotational
  sparse approximation. In particular, we aim to identify a rotation matrix
  such that the gPC expansion of a set of random variables after the rotation has a sparser representation. However, this rotational approach alters the underlying linear system to be solved, which
makes finding the sparse coefficients more difficult than the case without
  rotation. To solve this problem, we examine several popular nonconvex regularizations in compressive sensing (CS) that 
perform better than the classic $\ell_1$ approach empirically. All these regularizations can be minimized by the alternating direction method of multipliers (ADMM). 
Numerical examples show superior performance of the proposed combination of rotation and nonconvex sparse promoting regularizations over the ones without rotation and with rotation but using the convex $\ell_1$ approach.

  \noindent \textbf{keywords:}  
Generalized polynomial chaos, uncertainty quantification, iterative rotations,
  compressive sensing, alternating direction method of multipliers,
  nonconvex regularization .

\end{abstract}




\section{Introduction}
\label{sec:intro}
A surrogate model (also known as ``response surface'') plays an important role in uncertainty quantification (UQ), as it can efficiently evaluate the quantity of interest (QoI) of a system given a set of inputs. Specifically, in parametric uncertainty studies, the input usually refers to a set of parameters in the system, while the QoI can be observables such as mass, density, pressure, velocity, or even a trajectory of a dynamical system. The uncertainty in the system's parameters typically originates from the lack of physical knowledge, inaccurate measurements, etc. Therefore, it is common to treat these parameters as random variables, and statistics, e.g., mean, variance, and the probability density function (PDF) of the QoI with respect to such random parameters are crucial in understanding the behavior of the system. 

The generalized polynomial chaos (gPC) expansion~\cite{GhanemS91, XiuK02} is a widely used surrogate model in applied mathematics and engineering studies, which uses orthogonal polynomials associated with measures of the aforementioned random variables. 
Under some conditions, the gPC expansion converges to the QoI in a Hilbert space  as the number of polynomials increases~\cite{CameronM1947,ernst2012convergence, Ogura1972,XiuK02}.
Both \emph{intrusive} methods (e.g., stochastic Galerkin) and  \emph{non-intrusive} methods (e.g., probabilistic collocation method) \cite{BabuskaNT10,GhanemS91,TatangPPM97, XiuH05,XiuK02} have been developed  to compute the gPC coefficients. 
The latter is particularly more desirable to study a complex system, as it does not require to modify the computational models or simulation codes. 
For example, the gPC coefficients can be calculated based on input samples and corresponding output using least squared fitting, probabilistic collocation method, etc.

However, in many practical problems, it is prohibitive to obtain a large amount
of output samples used in the non-intrusive methods,
because it is costly to measure the QoI in experiments or to conduct simulations using a complicated model.
Consequently, one shall consider an under-determined linear system (matrix), denoted as 
$\tensor\Psi$, of size $M\times N$ with 
 $M<N$ (or even $M\ll N$), where $M$ is the size of available output samples and $N$ is the number of basis functions used in the gPC expansion. 
When the solution to the under-determined system is sparse,  compressive sensing (CS) techniques \cite{BrucksteinDE09,candes2008restricted,CRT,Donoho2006} are effective.
Recent studies have shown some success in applying CS to UQ problems 
\cite{adcock2017compressed,DoostanO11,LeiYLK17,LeiYZLB15,PengHD15,SargsyanSNDRT14,XuZ14,YanGX12,YangK13}. 
For example, sampling 
strategies~\cite{AlemazkoorM17,hampton2015compressive,jakeman2017generalized,Rauhut2012} can improve the property of $\tensor\Psi$ to guarantee sparse recovery via the $\ell_1$ minimization. Computationally, the 
weighted $\ell_1$ minimization \cite{Adcock17,candes2008enhancing,Peng2014,RauhutW15,YangK13} assigns larger weights to smaller components (in magnitude) of the solution, and hence minimizing
the weighted $\ell_1$ norm leads to a sparser solution than the  vanilla $\ell_1$ minimization does.  Besides,
adaptive basis selection~\cite{AlemazkoorM17,Blatman2011,dai2008subspace,Hampton2018,JakemanES14} as well as dimension reduction techniques can be adopted to reduce the number of unknown variables~\cite{tsilifis2019compressive,yang2018sliced} thus improving  computational efficiency. 

In this paper, we focus on a sparsity-enhancing approach, referred to as iterative rotation~\cite{LeiYZLB15,YangLBL16,yang2018atomic, yang2019general}, that intrinsically changes the structure of a surrogate model to make the gPC coefficients more sparse. However,  this method tends to deteriorate  properties of $\tensor\Psi$ that are favorable by CS algorithms, e.g., low coherence, which may counteract the benefit of the enhanced sparsity. 
Since the polynomials in the gPC expansion
may not be orthogonal after the rotation of the random variables,
the coherence of $\tensor\Psi$ may not converges to zero asymptotically, leading to an amplified coherence after the rotation.
To remedy this drawback, we innovatively combine the iterative rotation technique with
a class of nonconvex regularizations to improve the efficiency of CS-based UQ methods. Specifically, our new approach uses rotations to increase the sparsity while taking advantages of the nonconvex formalism for dealing with a matrix $\tensor\Psi$ that is highly coherent.
In this way, we leverage the advantages of both methods  to exploit information from limited samples of the QoI more efficiently and to construct gPC expansions more accurately. Main contributions of this work are two-fold. On one hand, we propose a unified and flexible framework that combines iterative rotation and sparse recovery together with an efficient algorithm. On the other hand, we empirically validate a rule of thumb in CS that nonconvex regularizations often lead to better performance compared to the convex approach.

The rest of the paper is organized as follows. We briefly review gPC, CS, and
rotational CS  in~\ref{sec:cs_uq}. We describe the combination of sparse signal
recovery and rotation matrix estimation in~\ref{sec:approach}.
\ref{sec:numeric} devotes to numerical examples, showing that the proposed approach significantly outperforms the state-of-the-art. 
Finally, conclusions are given in~\ref{sec:conclusion}.

\section{Prior works}
\label{sec:cs_uq}
In this section, we briefly review  gPC expansions,  useful concepts in CS, and
our previous work~\cite{YangLBL16,yang2019general} on the rotational CS for gPC methods.

\subsection{Generalized polynomial chaos expansions} 
We consider a QoI $u$ that depends on location $\bm x$, time $t$ and a set of random variables $\bx$ with the following 
 gPC expansion:
\begin{equation}\label{eq:gpc0}
u(\bm x, t;\bx) = \sum_{n=1}^Nc_n(\bm x, t)\psi_n(\bx) + \ve(\bm x, t;\bx),
\end{equation}
where $c_n(\bm x, t)= \mathbb{E}\{u(\bm x, t;\bm\xi)\psi_n(\bm\xi)\}$
and $\ve$ denotes the truncation error.
Here, $\{\psi_n\}_{n=1}^N$ are orthonormal with respect to the measure of $\bx$, i.e.,
\begin{equation}
  \int_{\mr^d} \psi_i(\bm x)\psi_j(\bm x)\rho_{\bx}(\bm x)\dif\bm x = \delta_{ij},
\end{equation}
where $\rho_{\bx}(\bm x)$ is the PDF of $\bx$, 
$\delta_{ij}$ is the Kronecker delta function, and we usually set
$\psi_1(\bm x)\equiv 1$.
We study systems relying on $d$-dimensional i.i.d.~random
variables $\bx=(\xi_1,\dotsc,\xi_d)$ and the gPC basis
functions are constructed by tensor products of univariate orthonormal 
polynomials associated with $\xi_i$. Specifically for a multi-index 
$\ba=(\alpha_1,\dotsc,\alpha_d)$ with each $\alpha_i\in\mn\cup\{0\}$, we set
\begin{equation}\label{eq:tensor}
\psi_{\ba}(\bx) =
\psi_{\alpha_1}(\xi_1)\psi_{\alpha_2}(\xi_2)\cdots\psi_{\alpha_d}(\xi_d),
\end{equation}
where $\psi_{\alpha_i}$ are univariate orthonormal polynomial of degree $\alpha_i$.
For two different multi-indices
$\bm{\alpha}=(\alpha_{_1}, \cdots, \alpha_{_d})$ and 
$\bm{\beta}=(\beta_{_1}, \cdots, \beta_{_d})$, we have 
\begin{equation}
  \int_{\mr^d} \psi_{\bm\alpha}(\bm x)\psi_{\bm\beta}(\bm x) \rho_{\bx}(\bm x) \dif\bm x=
\delta_{\bm\alpha\bm\beta} = \delta_{\alpha_1\beta_1}
\delta_{\alpha_2\beta_2}\cdots
\delta_{\alpha_d\beta_d},
\end{equation}
with $  \rho_{\bx}(\bm x) = \rho_{\xi_1}(x_1)\rho_{\xi_2}(x_2)\cdots\rho_{\xi_d}(x_d).$ 
Typically, a $p$th order gPC expansion involves all polynomials $\psi_{\bm\alpha}$ satisfying $|\bm\alpha|\leq p$ for $|\bm\alpha|=\sum_{i=1}^d \alpha_i$, which   indicates that a total number of $N=\left(\begin{smallmatrix}p+d \\ d\end{smallmatrix}\right)$ polynomials are used in the expansion.
For simplicity, we reorder $\bm\alpha$ in such a way that we index $\psi_{\bm\alpha}$ by $\psi_n$, which is consistent with Eq.~\eqref{eq:gpc0}.

In this paper, we focus on time independent problems, 
in which the gPC expansion at a fixed location $\bm x$ is given by
\begin{equation}
u(\bx^q) = \sum_{n=1}^N c_n\psi_n(\bx^q) + \ve(\bx^q), \quad q=1, 2, \cdots, M,
\end{equation}
where each $\bm \xi^q$ is a sample of $\bm\xi$, e.g, $\bm\xi^1=(\xi_1^1, \cdots,
\xi_d^1)$, and $M$ is the number of samples.
We rewrite the above expansion in terms of the matrix-vector notation, i.e.,
\begin{equation}\label{eq:cs_eq}
\tensor\Psi \bm c = \bm u - \bm\ve,
\end{equation}
where $\bm u=(u^1,\dotsc,u^M)\trans$ is a vector of output samples, 
$\bm c=(c_1, \dotsc, c_N)\trans$ is a vector of gPC coefficients,  $\tensor\Psi$ is an $M\times N$ matrix with $\Psi_{ij}=\psi_j(\bx^i),$ and 
$\bm\ve=(\ve^1,\dotsc,\ve^M)\trans$ is a vector of errors with 
$\ve^q=\ve(\bx^q)$.  
We are interested in identifying sparse coefficients $\h c= (c_1, \dotsc, c_N)\trans$ among the solutions of an under-determined system with $M<N$ in \eqref{eq:cs_eq},  which is the focus of compressive sensing \cite{candes2008introduction,donoho2006compressed,foucart2013mathematical}.  

\subsection{Compressive sensing}\label{sect:cs}

We review the concept of \emph{sparsity}, which plays an important role in  error
estimation for solving the under-determined system \eqref{eq:cs_eq}. The number of non-zero entries of a vector $\bm c=(c_1,\dotsc,c_N)$ is denoted by
$\Vert \bm c\Vert_0$.
Note that $\Vert\cdot\Vert_0$ is named the ``$\ell_0$ norm" in \cite{donoho2006compressed},
although it is not a norm nor a semi-norm. The vector $\bm c$ is called 
\emph{$s$-sparse} if $\Vert \bm c\Vert_0\leq s$, and it is considered a sparse vector if $s\ll N$. Few practical systems have truly 
sparse gPC coefficients, but rather \textit{compressible}, i.e., only a few
entries contributing significantly to its
$\ell_1$ norm. To this end, 
a vector $\bm c_s$ is defined as the best $s$-sparse approximation
which is obtained by setting all but the $s$-largest entries in magnitude of $\bm c$ to zero, and subsequently, $\bm c$ is regarded as sparse or compressible if 
$\Vert \bm c - \bm c_s\Vert_1$ is small for $s\ll N$. 

In order to find a sparse vector $\bm c$ from \eqref{eq:cs_eq}, one formulates the following  problem,
\begin{equation}\label{eq:l0}
  \hat{\bm c}_0=\arg\min_{\bm c}\frac 1 2 \Vert \tensor\Psi \bm c - \bm u\Vert_2^2 + \lambda\Vert \bm c\Vert_0,  
\end{equation}
where $\lambda$ is a positive parameter to be tuned such that $\Vert\tensor\Psi
\hat{\bm c}_0-\bm u\Vert_2\leq\epsilon.$ As the $\ell_0$ minimization \eqref{eq:l0} is NP-hard to solve \cite{natarajan95}, one often uses the convex  $\ell_1$ norm to replace 
 $\ell_0$, i.e.,
\begin{equation}\label{eq:l1}
  \hat{\bm c}_1=\arg\min_{\bm c}\frac 1 2 \Vert \tensor\Psi \bm c - \bm u\Vert_2^2 + \lambda\Vert \bm c\Vert_1.  
\end{equation} 
  A sufficient condition of the $\ell_1$ minimization to exactly recover the sparse signal was proved based on 
the \textit{restricted isometry property} (RIP) \cite{CRT}. Unfortunately,  RIP  is numerically unverifiable  for a given matrix
	\cite{bandeira2013certifying,tillmann2014computational}. Instead, a computable condition for $\ell_1$'s exact recovery is  \textit{coherence}, which is defined as 
\begin{equation}\label{eq:coherence}
\mu(\tensor\Psi) = \max_{i\neq j} \dfrac{|\langle \bm \psi_i,\bm \psi_j\rangle|}{\| \bm\psi_i\|\| \bm\psi_j\|}, \quad \mbox{with} \ \tensor\Psi=[ \bm\psi_1, \cdots, \bm \psi_{N}].
\end{equation} 
Donoho-Elad \cite{donohoE03} and Gribonval \cite{gribonval2003sparse} proved independently that if 
\begin{equation}\label{eq:coherence criterion}
\|\hat{\bm c}_1 \|_0<\frac 1 2 \left(1+\frac 1 {\mu(\tensor\Psi)}\right),
\end{equation}
then $\hat{\bm c}_1$ is indeed the sparsest solution to  \eqref{eq:l1}.
Although the inequality condition in \eqref{eq:coherence criterion} is not
sharp, the coherence of a matrix $\tensor\Psi$ is often used as an indicator to
quantify how difficult it is to find a sparse vector from a linear system
governed by $\tensor\Psi$ in the sense that the larger the coherence is, the
more challenging it is to find the sparse vector $\bm c$~\cite{louYHX14,yinLHX14}. 
Apparently if the samples $\bm\xi^q$ are drawn independently according to the
distribution of $\bm\xi$, $\mu(\tensor\Psi)$ for the gPC expansion converges to
zero as $M\rightarrow\infty$~\cite{DoostanO11}. However, the rotation technique 
tends to increase the coherence of the matrix, leading to unsatisfactory 
performance of the subsequent $\ell_1$ minimization as discussed in~\cite{yang2019general}.

In this work, we promote the use of nonconvex regularizations to find a
sparse vector when the coherence of $\tensor\Psi$ is relatively large, referred to as  a coherent linear system.
There are many nonconvex alternatives to approximate the $\ell_0$ norm that give
superior results over the $\ell_1$ norm, such as $\ell_{1/2}$
\cite{chartrand07,xuCXZ12,laiXY13}, capped $\ell_1$
\cite{zhang2009multi,shen2012likelihood,louYX16}, transformed
$\ell_1$~\cite{lv2009unified,zhangX17,zhangX18, GuoLLtransform18}, $\ell_1$-$\ell_2$  \cite{yinEX14,louYHX14,louY18},  $\ell_1/\ell_2$ \cite{l1dl2,L1dL2_accelerated} and error function (ERF) \cite{guo2020novel}.
We  formulate a general framework  that works for any  regularization whose proximal operator can be found efficiently. Recall that a proximal operator $\mathbf{prox}_{J}(\cdot)$ of a functional $J(\cdot)$ is defined by
\begin{equation}\label{eq:prox-def}
\mathbf{prox}_{J} (\bm c; \mu) \in \argmin_{y} \Big(\mu J(\bm y)+\frac 1 2 \|\bm y-\bm c\|_{2}^{2}\Big),
\end{equation}
where  $\mu$ is a positive parameter. 
We provide the formula of aforementioned regularizations together with their proximal operators as follows,
\begin{itemize}
\item The $\ell_{1}$ norm of $\h y$ is   $\|\bm y\|_1$ with its proximal operator given by
\begin{align}
 \label{eq:l_1 prox} \mathbf{prox}_{\ell_1}(\bm y;\mu) = \sign(\bm y)\circ \max(|\bm y|-\mu,0), 
\end{align}
where $\circ$ denotes the Hadamard operator for componentwise operation.
\item The square of the $\ell_{1/2}$ norm is defined as $\|\bm y \|_{1/2}^2 = \sum_n \sqrt{|y _{n}|},$ whose proximal operator \cite{xuCXZ12} is expressed as
\begin{equation}
   \label{eqn:l_p prox}\mathbf{prox}_{\ell_{1/2}}(\bm y;\mu) = \frac{3\h y}{4}\circ\left[\cos\left(\frac{\pi}{3} - \frac{\phi(\bm y)}{3}\right)\right]^2 \circ \max(\h y - \frac{3}{4}\mu^{2/3},0),
   \end{equation}
where $\phi(\bm y) = \displaystyle\arccos\left({\frac{\mu }{8} \big(\frac{3}{|\bm y|}\big)^{3/2}}\right)$ and the square is also computed componentwisely.
\item Transformed $\ell_1$ (TL1) is defined as $\frac{(\gamma+1)\| \bm y \|_{1}}{\gamma + \| \bm y \|_{1}}$ for a positive parameter $\gamma$ and its proximal operator \cite{zhangX17} is given by
\begin{align}
\label{eqn:tl1 prox}&\mathbf{prox}_{\mbox{TL1}}(\bm y;\mu) =
\begin{cases}
\left[\displaystyle\frac{2}{3}(\gamma +|\bm y|)\cos \frac{\phi(\bm y)}{3} -\frac{2}{3}\gamma + \frac{|\bm y|}{3}\right]
& \text{if } |\bm y|> \theta 
\\
0 &  \text{if } |\bm y| \le \theta,
\end{cases}    
\end{align}
with 
\begin{align*}
\phi(\bm y) = \displaystyle\arccos{\left(1 - \frac{27\mu \gamma(\gamma+1)}{2(\gamma + |\bm y|)^3}\right)} \quad \mbox{and}\quad \theta = 
\begin{cases}
 \mu\frac{\gamma+1}{\gamma}, & \text{if } \mu \le \frac{\gamma^2}{2(\gamma+1)} \\
\sqrt{2\mu(\gamma+1)} - \frac{\gamma}{2}, &  \text{if } \mu > \frac{\gamma^2}{2(\gamma+1)}.
\end{cases}
\end{align*}
\item The $\ell_{1}$-$\ell_{2}$ regularization is defined by $\|\bm y \|_{1} - \|\h y \|_{2},$ and 
its proximal operator \cite{louY18} is given by the following cases,
\begin{itemize}
    \item If $\|\bm y \|_{\infty} > \mu$, one has $\mathbf{prox}_{\ell_{1-2}}(\bm y;\mu) = \frac{\bm z (\|\bm z\|_{2} + \mu)}{ \|\tensor z\|_{2} }$, where $\bm z = \mathbf{prox}_{\ell_1}(\bm y;\mu)$.
    \item If $\|\bm y \|_{\infty} \leq \mu$, $\bm c^{*}:=\mathbf{prox}_{\ell_{1-2}}(\bm y;\mu)$ is an optimal solution if and only if   $ c^{*}_{i} = 0$ for $|y_{i}| < \|\bm y \|_{\infty}$, $\|\bm c^{*}\|_{2} = \|\bm y \|_{\infty},$ and $ c^{*}_{i}y_{i} \geq 0$ for all $i$. The optimality condition implies infinitely many solutions of $\bm c^*,$ among which we choose $c^{*}_{i} = \sign(y_{i}) \|y\|_{\infty}$ for the smallest $i$ satisfies $|y_{i}| = \|\bm y \|_{\infty}$ and the rest coefficients set to be zero.
\end{itemize}
\item The error function (ERF) \cite{guo2020novel} is defined by
\begin{equation}\label{eqn:erf}
J^{\mathrm{ERF}}_{\sigma}(\bm y) :=  \sum_{j=1}^n\int_0^{|y_j|}e^{-\tau^2/\sigma^2}d\tau,
\end{equation}
for $\sigma>0$. 
Though there is no closed-form solution for this problem, one can find the solution via the Newton's method. In particular, the optimality condition of \eqref{eq:prox-def} for ERF
 reads as
\begin{equation*}\label{eq:prox-optimality_ERF}
\h v\in \mu \partial J_\sigma^{\mathrm{ERF}}(\bm y) + \bm y = \mu \exp\left(-\frac{\bm y^2}{\sigma^2}\right)\odot\partial |\bm y| + \bm y.
\end{equation*}
When $|v_i|\leq \mu$, we have $x_i=0$. 
Otherwise the optimality condition becomes
\begin{equation*}
v_i= \mu \exp\left(-\frac{y_i^2}{\sigma^2}\right)\sign(v_i) + y_i,
\end{equation*}
which can be found by the Newton's method.
\end{itemize}


\subsection{Rotational compressive sensing}
\label{subsec:basic}
To further enhance the sparsity, we aim to find a linear map $\tensor A:\mr^d\mapsto\mr^d$ such that a new 
set of random variables $\bm\eta,$ given by 
\begin{equation}
  \bm\eta=\tensor A\tensor\bx, \quad \bm\eta =
  (\eta_1,\eta_2,\cdots,\eta_d)\trans,
\end{equation}
leads to a sparser polynomial expansion than $\bx$ does.
We consider $\tensor A$ as an orthogonal matrix, i.e., $\tensor A\tensor A\trans=\tensor I$ with the identity matrix $\tensor I$, such that the linear map
from $\bx$ to $\bm\eta$ can be regarded as a rotation in $\mathbb{R}^d$.
Therefore, the new polynomial expansion for $u$ is expressed as
\begin{equation}\label{eq:new_poly}
u(\bx)\approx u_g(\bx) =
\sum_{n=1}^N\tilde c_n\psi_n(\tensor A\bx)=
\sum_{n=1}^N\tilde c_n\psi_n(\bm\eta) = v_g(\bm\eta).
\end{equation}
Here $u_g(\bx)$ can be understood as a polynomial $u_g(\bm x)$ evaluated at random 
variables $\bx$, and the same for $v_g$. Ideally, $\tilde{\bm c}$ is sparser than 
$\bm c$. 
In the previous works \cite{LeiYZLB15,YangLBL16}, it is assumed that 
$\bm\xi\sim\mathcal{N}(\bm 0, \tensor I)$, so 
$\bm\eta\sim\mathcal{N}(\bm 0, \tensor I)$. For general cases where
$\{\xi_i\}_{i=1}^d$ are not i.i.d.~Gaussian, $\{\eta_i\}_{i=1}^d$ are not 
necessarily independent. Moreover, $\{\psi_n\}_{n=1}^N$ are not necessarily 
orthogonal to each other with respect to $\rho_{\bm\eta}$. Therefore,
$v_g(\bm\eta)$ may not be a standard 
gPC expansion of $v(\bm\eta)$, but rather
a polynomial equivalent to $u_g(\bm\xi)$ with potentially sparser coefficients~\cite{yang2019general}.


We can identify $\tensor A$ using the gradient information of $u$ based on the framework of active subspace~\cite{ConstantineDW14,Russi10}. In particular, we define
\begin{equation}\label{eq:grad_u}
  \tensor W=\dfrac{1}{\sqrt{M}}[\nabla u(\bm\xi^1),\nabla u(\bm\xi^2),\cdots,\nabla u(\bm\xi^{M})].
\end{equation}
Note that $\tensor W$ is a $d\times M$ matrix, and we consider $M\geq d$ in this work. The singular value decomposition
(SVD) of $\tensor W$ yields
\begin{equation}\label{eq:pca}
\tensor W = \tensor U \tensor \Sigma\tensor V\trans,
\end{equation}
where $\tensor U$ is a $d\times d$ orthogonal matrix, 
$\tensor \Sigma$ is a $d\times M$ matrix, whose diagonal consists of singular
values $\sigma_1\geq\cdots\geq\sigma_d\geq 0$, and $\tensor V$ is an $M\times M$ orthogonal matrix. We set the rotation matrix as
$\tensor A = \tensor U\trans$. As such, the rotation projects $\bm\xi$ to the
directions of principle components of $\nabla u$. Of note, we do not use the
information of the orthogonal matrix $\tensor V$. 

Unfortunately, $u$ is unknown, and thus the samples of $\nabla u$ are not
available. Instead, we approximate Eq.~\eqref{eq:grad_u} by a computed solution $u_g$ that can be obtained by the $\ell_1$ minimization 
\cite{YangLBL16}. In other words, we have
\begin{equation}
  \tensor W\approx\tensor W_g=\dfrac{1}{\sqrt{M}}[\nabla u_g(\bm\xi^1),\nabla
u_g(\bm\xi^2),\cdots,\nabla u_g(\bm\xi^{M})],
\end{equation}
and the rotation matrix is constructed based on the SVD of $\tensor W_g$:
\begin{equation}\label{eq:pcag}
\tensor W_g = \tensor U_g \tensor \Sigma_g\tensor V\trans_g, \quad
\tensor A = \tensor U_g\trans.
\end{equation}
Defining $\bm\eta=\tensor A\bm\xi,$ we compute the corresponding input samples
as $\bm\eta^q=\tensor A\bm\xi^q$ and construct a new measurement 
matrix $\tensor\Psi(\bm\eta)$ as $(\tensor\Psi(\bm\eta))_{ij}=\psi_j(\bm\eta^i)$.
We then solve the minimization problem in Eq.~\eqref{eq:l1} to obtain
$\tilde{\bm c}$. 
If some singular values of $\tensor W_g$ are much larger than the others, we can expect to obtain a sparser representation of $u$ with respect to
$\bm\eta,$ which is dominated by the eigenspace associated with these larger singular
values. 
On the other hand, if all the singular values $\sigma_i$ are of the same order, the rotation does not enhance the sparsity.
In practice, this method can be designed as an iterative algorithm, in which $\tensor A$ and $\tilde{\bm c}$ are updated separately  following an alternating direction manner. 

It is worth noting that the idea of using a linear map to identify a possible
low-dimensional structure is also used in sliced inverse 
regression (SIR) \cite{Li1991}, active subspace \cite{ConstantineDW14,Russi10},
basis adaptation \cite{TipG14}, etc., but with
different manners in computing the matrix. In contrast to these methods, the iterative rotation approach does
not truncate the dimension in the sense that $\tensor A$ is a square matrix. As an initial guess may not be sufficiently accurate, reducing dimension before the iterations terminate
may lead to suboptimal results.
The dimension reduction was integrated 
with an iterative method in \cite{yang2017pdf}, while
another iterative rotation method preceded with 
SIR-based dimension reduction was proposed in \cite{yang2018sliced}. We refer
interested readers to the respective literature.

\section{The proposed approach}
\label{sec:approach}
When applying the rotational CS techniques, the measurement matrix $\tensor\Psi(\bm\eta)$ may become more coherent compared with $\tensor\Psi$. This is because popular polynomials $\psi_i$ used in gPC method, e.g., Legendre and Laguerre polynomials, are not orthogonal with respect to the measure of $\bm\eta$, so $\mu(\tensor\Psi(\bm\eta))$ converges to a positive number instead of zero as $\mu(\tensor\Psi)$ does.
Under such a coherent regime, we  advocate the minimization of a nonconvex  regularization to identify the sparse coefficients. 

To start with, we generate input samples $\{\bx^q\}_{q=1}^M$ based on the distribution
     of $\bx$ and select the gPC basis functions $\{\psi_j\}_{j=1}^N$ associated with $\bx$ in order to generate the measurement matrix $\tensor\Psi$ by setting
  $\Psi_{ij}=\psi_j(\bx^i)$ as in Eq.~\eqref{eq:cs_eq}, while initializing $\tensor A^{(0)}=\tensor I$ and
$\bm\eta^{(0)}=\bm\xi.$ Then we propose an alternating direction method (ADM) that combines the nonconvex minimization  and rotation matrix estimation. Specifically given $\{\bx^q\}_{q=1}^M,$ $\{\psi_j\}_{j=1}^N,$ and $\bm u:=\{u^q=u(\bx^q)\}_{q=1}^M,$ we formulate the following minimization problem,
\begin{equation}\label{eq:rot_l1l2}
\argmin_{\bm c,~ \tensor A} \lambda J(\bm c)+\frac 1 2\Vert\tensor \Psi\bm c-\bm u\Vert_2^2,
  \quad \tensor A\tensor A\trans=\tensor I \mbox{ and } \tensor\Psi_{ij}=\psi_j(\tensor A\bm\xi^i),
\end{equation}
where $J(\cdot)$ denotes a regularization functional and $\lambda>0$ is a weighting parameter. For example, our early work \cite{YangLBL16} considered  $J(\bm c)=\|\bm c\|_1$, i.e., the $\ell_1$ approach.
We can minimize Eq.~\eqref{eq:rot_l1l2} with respect to $\bm c$ and $\tensor A$ in an alternating direction manner. 
When $\tensor A$ is fixed, we minimize a sparsity-promoting functional $J(\cdot)$ to identify the sparse coefficients $\bm c,$ as detailed in Section~\ref{sect:l1-l2}. 
When $\bm c$ is fixed, optimizing $\tensor A$ is computationally expensive, 
unless a dimension reduction technique is used, e.g., as in~\cite{tsilifis2019compressive}. Instead, we use the rotation estimation introduced in Section~\ref{subsec:basic} to construct $\tensor A$. 
Admittedly, this way of estimating $\tensor A$ may not be optimal, but it can
  potentially promote the sparsity of $\bm c$, thus improving the accuracy of the sparse approximation of the gPC expansion.

\subsection{Finding sparse coefficient via ADMM}\label{sect:l1-l2}

We focus on the $\bm c$-subproblem in \eqref{eq:rot_l1l2}, whose objective function is defined as
\begin{align}\label{eq:general obj}
\tilde{\bm c}: =\argmin_{\bm c}\lambda J(\bm c)+\frac 1 2\Vert\tensor \Psi\bm c-\bm u\Vert_2^2.
\end{align}
We adopt  the alternating direction method of multipliers (ADMM) \cite{boydPCPE11admm} to minimize \eqref{eq:general obj}.
In particular, we introduce an auxiliary variable $\h y$ and rewrite \eqref{eq:general obj} as an equivalent problem,
\begin{equation}\label{eq:L1uncon_split}
\min_{\bm c,\bm y} \lambda J(\bm c) + \frac 1 2\|\tensor\Psi\bm y-\bm u\|_2^2  \quad \mbox{s.t.} \quad \bm c=\bm y.
\end{equation}
This new formulation \eqref{eq:L1uncon_split} makes the objective function separable with respect to two variables $\bm c$ and $\bm y$ to enable efficient computation.  Specifically, the  augmented Lagrangian  corresponding to \eqref{eq:L1uncon_split} can be expressed as
	\begin{equation}\label{Lagrangian l_1-2 }
    L_\rho(\bm c,\bm y;\bm w) = \lambda J(\bm c)  + \frac 1 2\|\tensor\Psi\bm y-\bm u\|_2^2 + \langle \bm w, \bm c-\bm y\rangle+\frac \rho 2\|\bm c-\bm y\|_2^2,
	\end{equation}
	where $\bm w$ is an Lagrangian multiplier and $\rho$ is a positive parameter.
Then the ADMM iterations indexed by $k$ consist of three steps,
\begin{eqnarray}
\label{eq:ADMM-x} \bm c_{k+1} &=& \argmin_{\bm c} \lambda J(\bm c) +\frac \rho 2 \|\bm c-\bm y_{k}+\frac {\bm w_{k}} \rho\|_2^2\\
\label{eq:ADMM-y}\bm y_{k+1} &=&\argmin_{\bm y} \frac 1 2 \|\tensor\Psi\bm y-\bm u\|_2^2  +\frac \rho 2 \|\bm c_{k+1}-\bm y+\frac {\bm w_{k}} \rho\|_2^2\\
\label{eq:ADMM-lambda}\bm w_{k+1} &=& \bm w_{k}+\rho(\bm c_{k+1}-\bm y_{k+1}).
\end{eqnarray}
Depending on the choice of $J(\cdot),$ the $\bm c$-update  \eqref{eq:ADMM-x} can be given by its corresponding proximal operator, i.e.,
\begin{equation}
\bm c_{k+1} = \mathbf{prox}_{J} \left(\bm y_{k}-\frac {\bm w_{k}} \rho;\frac{\lambda}{\rho} \right).
\end{equation} 
Please refer to Section \ref{sect:cs} for  detailed formula of proximal operators.

As for the $\bm y$-update, we take the gradient of  \eqref{eq:ADMM-y} with respect to $\bm y$, thus leading to
\begin{equation}\tensor\Psi^{T}(\tensor\Psi\bm y-\bm u) + \rho(\bm y-\bm c_{k+1}-\frac {\bm w_{k}} \rho) =\bm 0.
\end{equation}
Therefore, the update for $\h y$ is given by 
\begin{eqnarray}
\bm y_{k+1} = (\tensor\Psi^{T} \tensor\Psi+\rho \tensor I)^{-1} \left(\tensor\Psi\trans\bm u   +\rho \bm c_{k+1}+\bm w_{k}\right).
\end{eqnarray}
Note that $\tensor\Psi\trans\tensor\Psi+\rho \tensor I$ is a positive definite matrix and there are many efficient numeral algorithms for matrix inversion. Since $\tensor\Psi$ has more columns than rows in our case, we further use the \textit{Woodbury formula} to speed up, 
\begin{eqnarray}
\rho(\tensor\Psi\trans\tensor\Psi+\rho \tensor I)^{-1}  = \tensor I-\frac 1 \rho \tensor\Psi\trans(\tensor\Psi\tensor\Psi\trans+\rho \tensor I)^{-1},
\end{eqnarray}
as $\tensor\Psi\tensor\Psi\trans$ has a smaller dimension than  $\tensor\Psi\trans\tensor\Psi$ to be inverted. In summary, the overall minimization algorithm based on ADMM  is described in~\ref{algo:cs_l1l2}.

\begin{algorithm}[t]
 \caption{The ADMM framework for solving a general sparse coding problem \eqref{eq:general obj}.}
\label{algo:cs_l1l2}
\begin{algorithmic}[1]
    \STATE Input: measurement matrix $\tensor\Psi$ and observed data $\bm u$.
    \STATE Parameters: $\lambda$, $\rho$, $\epsilon \in \mathbb{R^+}$ and $\text{k}_{\text{max}}\in \mathbb Z^+$. 
    \STATE Initialize: $\bm c,\bm y,\bm v, \bm w$ and $k = 0$.
    \WHILE {$k \leq \text{k}_{\text{max}}$ or $\| \bm c_{k} - \bm c_{k-1} \| > \epsilon$} 
        \STATE $\bm c_{k+1} =  \mathbf{prox}_{J} \left(\bm y_{k}-\frac {\bm w_{k}} \rho;\frac{\lambda}{\rho} \right)$
        \STATE $\bm y_{k+1} = (\tensor\Psi\trans\tensor\Psi+\rho \tensor I)^{-1} (\tensor\Psi\trans\bm u   + \rho \bm c_{k+1}+\bm w_{k})$
        \STATE $\bm w_{k+1} = \bm w_{k}+\rho(\bm c_{k+1}-\bm y_{k+1})$
        \STATE $k = k+1$
    \ENDWHILE
    \STATE Return $\tilde{\bm c} = \bm c_{k}$
\end{algorithmic}
\end{algorithm}

\subsection{Rotation matrix update}\label{sect:rot}
Suppose we obtain the gPC
coefficients $\tc^{(l)}$  at the $l$-th iteration via Algorithm~\ref{algo:cs_l1l2} with $l\geq 1$. Given $v_g^{(l)}(\bm\eta)$ with input samples
$\{(\bm\eta^{(l)})^q\}_{q=1}^M$ for $(\bm\eta^{(l)})^q=\tensor A^{(l-1)}(\bm\eta^{(l-1)})^q$, we collect the gradient of $v_g^{(l)}$, denoted by
\begin{equation}\label{eq:grad_iter}
  \tensor W_g^{(l)}=\dfrac{1}{\sqrt{M}}\left [\nabla_{\bm\xi} v_g^{(l)}\left((\bm\eta^{(l)})^1\right),
    \cdots, \nabla_{\bm\xi} v_g^{(l)}\left((\bm\eta^{(l)})^{M}\right)\right ],
\end{equation}
where $\nabla_{\bm\xi}\cdot=(\partial\cdot/\partial \xi_1, \partial\cdot/\partial \xi_2,
\cdots, \partial\cdot/\partial \xi_d)\trans$. 
It is straightforward to evaluate $\nabla\psi_n$ at $(\bm\eta^{(l)})^q,$ as we construct $\psi_n$ using the tensor product of
univariate polynomials \eqref{eq:tensor} and derivatives for widely used
orthogonal polynomials, e.g., Hermite, Laguerre, Legendre, and
Chebyshev, are well studied in the UQ literature. Here, we can analytically 
compute the gradient of $v_g^{(l)}$ with respect to $\bx$ using the chain rule, 
\begin{equation}\label{eq:chain}
  \begin{split}
 &  \nabla_{_{\bm\xi}} v_g^{(l)}\left((\bm\eta^{(l)})^q\right)=
  \nabla_{_{\bm\xi}} v_g^{(l)}\left(\tensor A^{(l)}\bm\xi^q\right) 
  = (\tensor A^{(l)})\trans \nabla v_g^{(l)}(\bm x) \bigg |_{\bm x=\tensor A^{(l)}\bm\xi^q} \\
  = &  (\tensor A^{(l)})\trans\nabla \sum_{n=1}^N \tilde c_n^{(l)}\psi_n(\bm x)\bigg |_{\bm x=\tensor A^{(l)}\bm\xi^q}
  = (\tensor A^{(l)})\trans \sum_{n=1}^N \tilde c_n^{(l)}\nabla \psi_n(\bm x)\bigg |_{\bm x=\tensor A^{(l)}\bm\xi^q} .
  \end{split}
\end{equation}
Then, we have an update for $\tensor A^{(l+1)}=\left(\tensor U_g^{(l)}\right)\trans,$ where $\tensor U_{W}^{(l)}$ is from  the SVD of 
\begin{equation}\label{eq:svd_iter}
\tensor W_g^{(l)} = \tensor U_g^{(l)} \tensor
\Sigma_g^{(l)}\left(\tensor V_g^{(l)}\right)\trans.
\end{equation}
Now we can define a new set of random variables as $\bm\eta^{(l+1)}=\tensor A^{(l+1)}\bm\xi$ and
compute their samples accordingly: $(\bm\eta^{(l+1)})^q=\tensor A^{(l+1)}\bm\xi^q$. 
These samples are then used to construct a new measurement matrix $\tensor\Psi^{(l+1)}$ as 
$\Psi^{(l+1)}_{ij}=\psi_j((\bm\eta^{(l+1)})^i)$ to feed into~\ref{algo:cs_l1l2} to obtain $\bm c^{(l+1)}$. 

We summarize the entire procedure in~\ref{algo:cs_rot1}.
\begin{algorithm}[t]
 \caption{Alternating direction method of minimizing \eqref{eq:rot_l1l2}}
\label{algo:cs_rot1}
\begin{algorithmic}[1]
  \STATE Generate input samples $\{\bx^q\}_{q=1}^M$ based on the distribution
     of $\bx$. 
\STATE Generate corresponding output samples $\bm u:=\{u^q=u(\bx^q)\}_{q=1}^M$ by
    solving the complete model, e.g., running simulations, solvers, etc.
\STATE Select gPC basis functions $\{\psi_n\}_{n=1}^N$ associated with $\bx$ and set counter $l=0$. Set $\tensor A^{(0)}=\tensor I$, $\bm \tilde{c}^{(0)} = \bm 0$, $e = 1$ and $\bm\eta^{(0)}=\bm\xi$.
\STATE Generate the measurement matrix $\tensor\Psi^{(l)}$ by setting
$\Psi_{ij}^{(l)}=\psi_j(\bx^i)$.
\WHILE{$l<l_{\max}$ and $e \geq 10^{-3}$}
\IF {$l>0$}
\STATE Construct $\tensor W^{(l)}$ in \eqref{eq:grad_iter} with  $v_g^{(l)}=\sum_{n=1}^N\tilde c^{(l)}\psi_n(\bx^{(l)})$.
\STATE  Compute SVD of $\tensor W_g^{(l)}$:
$\tensor W_g^{(l)}= \tensor U^{(l)}_{g}\tensor\Sigma^{(l)}_{g}\left(\tensor
  V^{(l)}_{g}\right)\trans.$
\STATE Set $\tensor A^{(l+1)}=\left(\tensor U^{(l)}_{g}\right)\trans$ and 
  $\bm\eta^{(l+1)}=\tensor A^{(l+1)}\bm\xi$.
 \STATE Construct the new measurement matrix $\tensor\Psi^{(l+1)}$ with
$\Psi^{(l+1)}_{ij}=\psi_j\left((\bm\eta^{(l+1)})^i\right)$. 
\ENDIF
\STATE Solve the minimization problem via ADMM (Algorithm~\ref{algo:cs_l1l2}): 
\[\tc^{(l+1)}=\argmin_{\h c} \lambda J(\h c) +\frac 1 2\Vert\tensor \Psi^{(l+1)}\h c-\bm u\Vert_2^2.\]

\STATE Calculate root square mean error of coefficients between two consecutive iterations 
\[
    e = \frac{\|\tc^{(l+1)} - \tc^{(l)}\|_2}{\|\tc^{(l+1)}\|_2}.
\]
\STATE $l=l+1$. 
\ENDWHILE
\STATE Construct gPC expansion as  
$u(\bx)\approx u_g(\bm\xi)=v_g^{(l_{\max})}(\bm\eta^{(l_{\max})})=\sum\limits_{n=1}^N \tilde{c}^{(l_{\max})}_n\psi_n(\tensor A^{(l_{\max})}\bx)$.
\end{algorithmic}
\end{algorithm}
The stopping criteria we adopt are $l\leq l_{\max}$ and the relative difference of coefficients between two consecutive iterations less than $10^{-3}$. As the sparsity structure is problem-dependent (see examples in
Section~\ref{sec:numeric}), more iterations do not grant significant improvements for many practical problems. 

\section{Numerical Examples}\label{sec:numeric}
In this section, we present four numerical examples to
demonstrate the performance of the proposed method. Specifically, in
Section~\ref{sect:ridge}, we examine a function with low-dimensional structure,
which has a truly sparse representation. Section~\ref{sect:elliptic} discusses
an elliptic equation widely used in UQ literature
whose solution is not exactly sparse. The example discussed in
Section~\ref{sect:kdv} has two dominant directions, even though the solution is
not exactly sparse either. Lastly, we present a high-dimensional example in
Section~\ref{sect:HD}. These examples revisit some numerical tests in previous
works~\cite{YangLBL16, yang2019general}, and hence provide direct comparison of 
the newly proposed approaches and the original one.

We compare the proposed framework  to the rotation with the $\ell_{1}$ approach (by setting $J(\h c)=\|\h c\|_1$ in \eqref{eq:rot_l1l2}) and the one without rotation (by setting $l_{\max}=1$ in~\ref{algo:cs_rot1}). The performance is evaluated in terms of relative error (RE), defined as $\frac{\|u - u_g\|_2}{\|u \|_2}$,
where $u$ is the exact solution, $u_g$ is a reconstructed solution as a gPC approximation of $u$, and 
the integral in computing the norm $\Vert\cdot\Vert_2$ is approximated with a high-level sparse grids method, based on one-dimensional Gaussian quadrature and the Smolyak structure~\cite{Smolyak1963}. 

We set $l_{\max}=9$. 
To tune for other two parameters $(\lambda,\rho)$,
we generate $10$ independent random trials 
and start with a coarser grid of $10^{-4}$, $10^{-3}$, $\cdots$, $10^{1}$. For each combination of $\lambda$ and $\rho$ in this coarse grid, we apply \ref{algo:cs_rot1} for every regularization functional that is mentioned in Section~\ref{sect:cs} and record all the relative errors. 
After finding the optimal exponential order that achieves the smallest averaged RE over these random trials,  we multiply  by  $0.5$, $1$, $\cdots$, $9.5$ to find the parameter values, denoted by $(\bar\lambda,\bar\rho).$
We specify the parameter pair $(\bar\lambda,\bar\rho)$ for each testing
case in the corresponding section. 
After $(\bar\lambda, \bar\rho)$ are identified, we conduct another $100$
independent random trials (not including trials for the parameter tuning
procedure), and report the average RE.
In practice, these parameters can be determined by the $k$-fold 
cross-validation following the work of \cite{DoostanO11}. Empirically, we
did not observe significant difference between these two parameter tuning
procedure for our numerical experiments, so we used a fixed set of 
$(\lambda, \rho)$ for all $100$ trials in each example.

In the first example, we do not terminate the iteration when $e>10^{-3}$ in
Algorithm~\ref{algo:cs_rot1}, but finish $nine$ iterations to compare the
performance of different approaches when a truly sparse representation is
available after a proper rotation. In the remaining examples, we follow
the termination criterion (i.e., line 5\ in Algorithm~\ref{algo:cs_rot1}).

\subsection{Ridge function}\label{sect:ridge}
Consider the following ridge function:
\begin{align}
    u(\h{\xi})=\sum_{i=1}^{d} \xi_{i}+0.25\left(\sum_{i=1}^{d} \xi_{i}\right)^{2}+0.025\left(\sum_{i=1}^{d} \xi_{i}\right)^{3}.
\end{align}
As $\{\xi_{i}\}_{i=1}^d$ are equally important in this example, adaptive methods \cite{Ma2010,Yang2012,Zhang2015}  that build surrogate
models hierarchically based on the importance of $\xi_{i}$ may not be effective. We consider a rotation matrix in the form of
\begin{align}
    \mathbf{A}=\left(\begin{array}{cccc}
d^{-\frac{1}{2}} & d^{-\frac{1}{2}} & \cdots & d^{-\frac{1}{2}} \\
\\
& \tilde{\mathbf{A}} &\\
&&
\end{array}\right),
\end{align}
where $\tilde{\tensor A}$ is an $d \times (d-1)$ matrix designed to guarantee the orthogonality of the matrix $\tensor A$, e.g., $\tensor  A$ can be obtained by the Gram–Schmidt process. With this choice of  $\tensor A$, we have $\eta_{1} = d^{-\frac{1}{2}} \sum_{i=1}^{d} \xi_{i} $ and $u$ can be represented as
\begin{align}
    u(\bm{\xi})=v(\bm{\eta})=d^{\frac{1}{2}} \eta_{1}+0.25 d \eta_{1}^{2}+0.025 d^{\frac{3}{2}} \eta_{1}^{3}.
\end{align}
In the expression $u(\bm\xi)=\sum_{n=1}^N\tilde c_n\psi_n(\tensor A\bm\xi)=\sum_{n=1}^N\tilde c_n\psi_n(\bm\eta)$,
all of the polynomials that are not related to $\eta_{1}$ make no contribution to the expansion, which guarantees the sparsity of  $\tilde{\bm c}=(\tilde c_1,\dotsc,\tilde c_N)$. 


By setting $d = 12$ (hence, the number of gPC basis functions is $N = 455$ for $p = 3$), 
 we compare the accuracy of computing gPC expansions by minimizing different regularization functionals with and without rotations. We consider gPC expansion using Legendre polynomial (assuming $\xi_{i}$ are i.i.d.~uniform random variables), Hermite polynomial expansion (assuming $\xi_{i}$ are i.i.d.~Gaussian random variables), and Laguerre polynomial (assuming $\xi_{i}$ are i.i.d.~exponential random variables), respectively. 
 The number of sample $M$ ranges from $100$ to $180$, each repeated 100 times to compute the average RE.  
 The parameters $\bar\lambda,\bar\rho$ for all the regularization models are listed in \ref{tab:parameter ridge}.

\begin{table}[t]
    \centering
    \caption{Parameters $(\bar\lambda,\bar\rho)$ for different regularization models in the case of ridge function. } 
    \label{tab:parameter ridge}
    \begin{tabular}{c|c c| c c| c c}
    \hline  \hline
  UQ setting   & \multicolumn{2}{c|}{ Legendre }& \multicolumn{2}{c|}{Hermite}    &\multicolumn{2}{c}{Laguerre }    \\ 
       &$\bar\lambda$  &$\bar\rho$&$\bar\lambda$  &$\bar\rho$&$\bar\lambda$  &$\bar\rho$\\
    \hline
        $\ell_1$     & $6\times 10^{-4}$& $6\times 10^{-1}$     & $1\times 10^{-2}$& $1\times 10^{-1}$      & $5\times 10^{-2}$& $5\times 10^{0}$ \\
        $l_{1/2}$    & $5 \times 10^{-3}$& $ 1.6\times 10^{1}$    & $1.8 \times 10^{-2}$& $6 \times 10^{-1}$  & $6\times 10^{-2}$& $1.2\times 10^{0}$ \\
        TL$1$  & $1\times 10^{-8}$& $ 1\times 10^{-7}$ & $1\times 10^{-8}$& $ 1\times 10^{-7}$ & $1\times 10^{-8}$& $1\times 10^{-7}$ \\
        ERF          & $1.5\times 10^{-1}$& $1.6\times 10^{3}$   & $1 \times 10^{0}$& $1.2 \times 10^{1}$    & $1.1 \times 10^{0}$& $1.5 \times 10^{3}$ \\
        $\ell_1 - \ell_2$ & $3\times 10^{-4}$& $ 5\times 10^{-1}$ & $ 1\times 10^{-4}$& $1\times 10^{-2}$ & $5\times 10^{-1}$& $1\times 10^{0}$\\
    \hline
    \end{tabular}
\end{table}

\begin{figure}[t]
  \centering
  \subfigure[Legendre]{
    \includegraphics[width=0.3\textwidth]{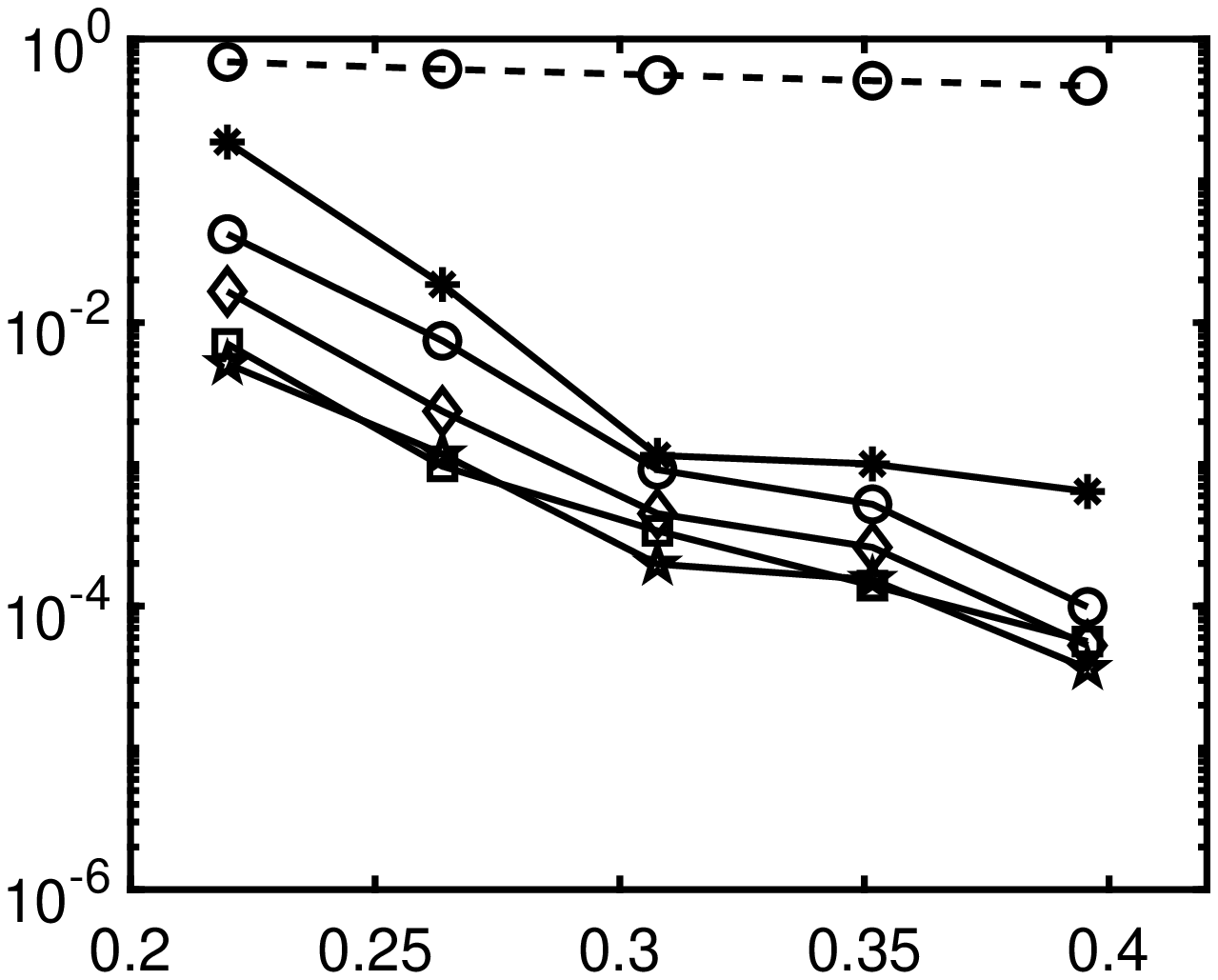}}  
    \subfigure[Hermite]{
     \includegraphics[width=0.3\textwidth]{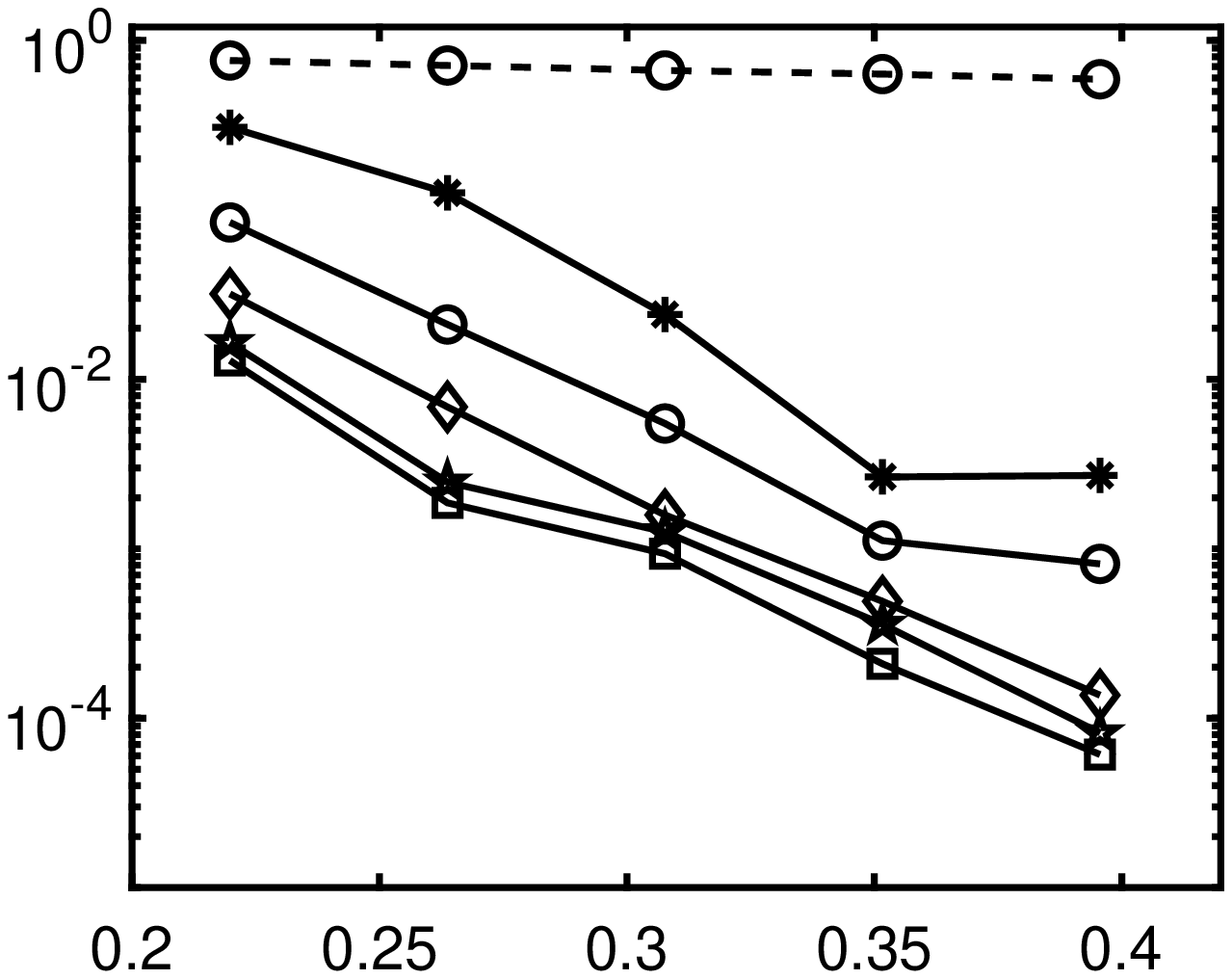}}
    \subfigure[Laguerre]{
    \includegraphics[width=0.3\textwidth]{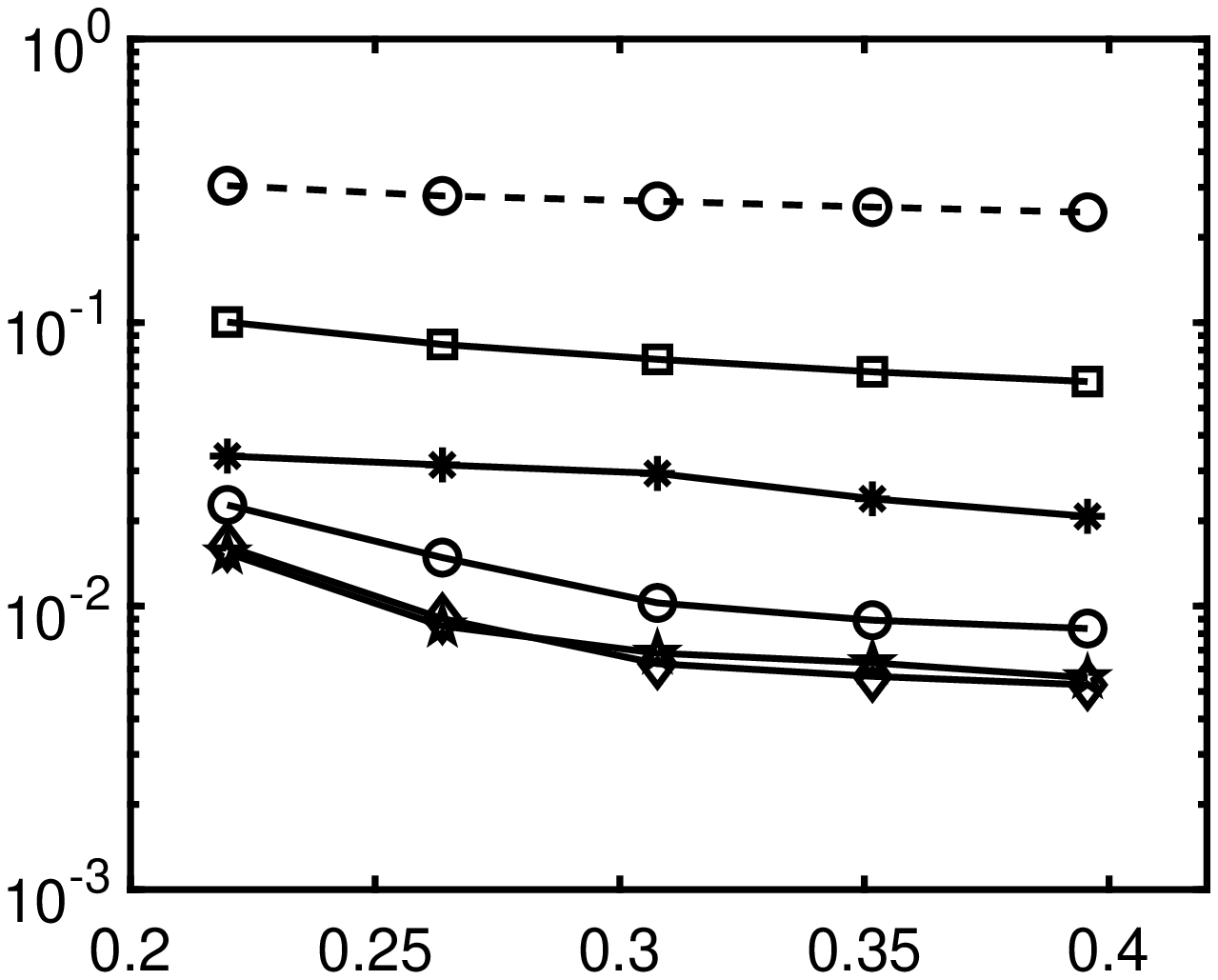} }
    \caption{Relative errors of Legendre, Hermite and Laguerre polynomial expansions for ridge function against ratio $M/N$.  Solid lines are for experiments with rotations applied whereas dashed line is a reference of $\ell_{1}$ test result without rotation. `` $\circ$'' is the marker for $\ell_{1}$ minimization, `` $*$'' is the marker for $\ell_{1/2}$, `` $\square$'' is the marker for TL1, `` $\star$'' is the marker for ERF, `` $\Diamond $'' is the marker for $\ell_{1} - \ell_{2}$.}
    \label{fig:ridge}
\end{figure}

\ref{fig:ridge} plots relative errors corresponding to the ratios of $M/N$, showing that nonconvex regularizations (except for $\ell_{1/2}$)  outperform the convex $\ell_1$ approach.
In particular, the best regularizations for different polynomials are different,
but $\ell_{1}$-$\ell_{2}$ and ERF generally perform very well (within top two). TL1 works well in the cases of Legendre and Hermite polynomials, while $L_{1/2}$ is not numerically stable to guarantee satisfactory results. Please refer to Section~\ref{sect:disc} for in-depth discussions on the performance of these regularizations with respect to the matrix coherence.

\begin{figure}[t]
  \centering
  \subfigure[Legendre $|c_n|$]{
    \includegraphics[width = 0.3\textwidth]{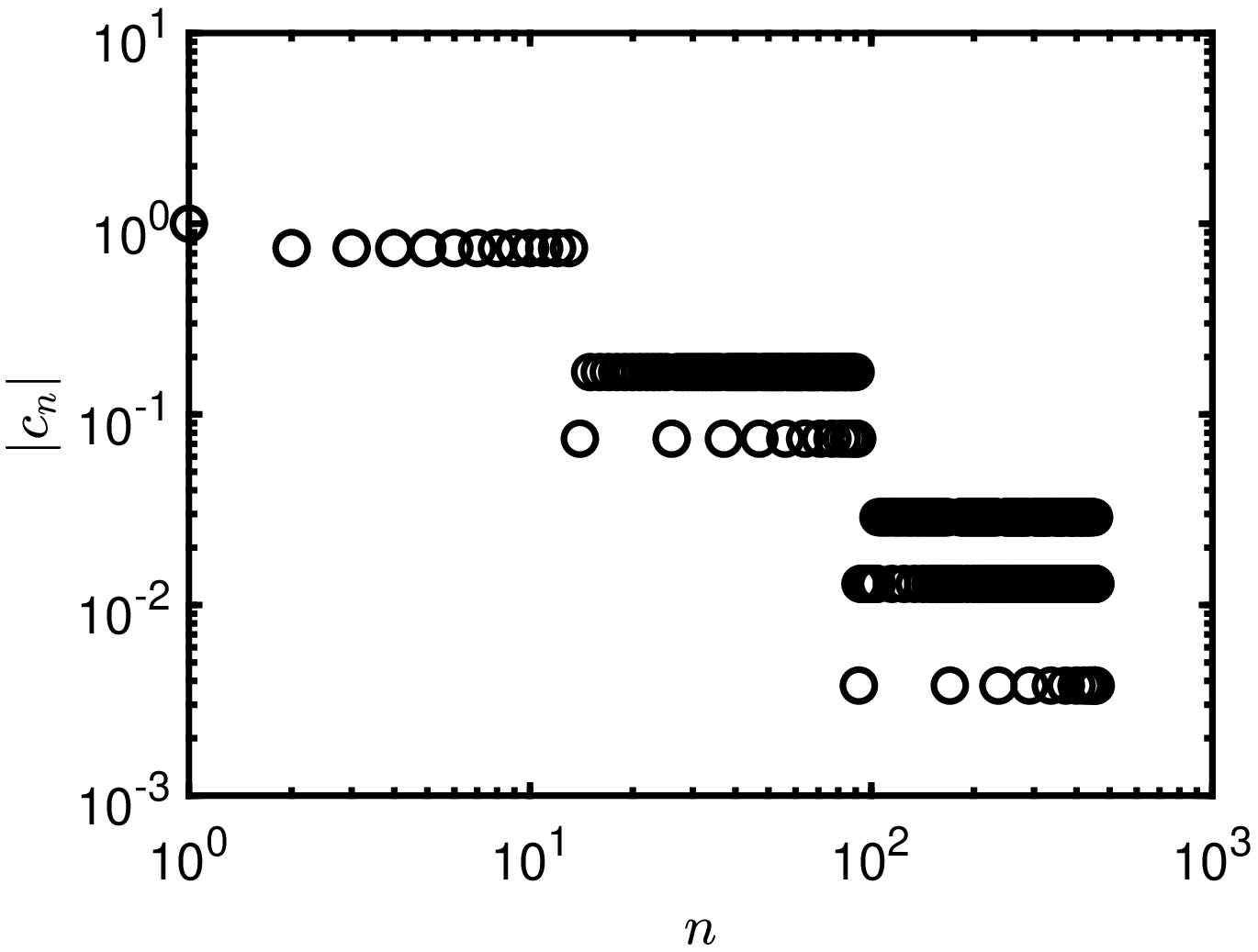}}
    \subfigure[Hermite $|c_n|$]{
    \includegraphics[width = 0.3\textwidth]{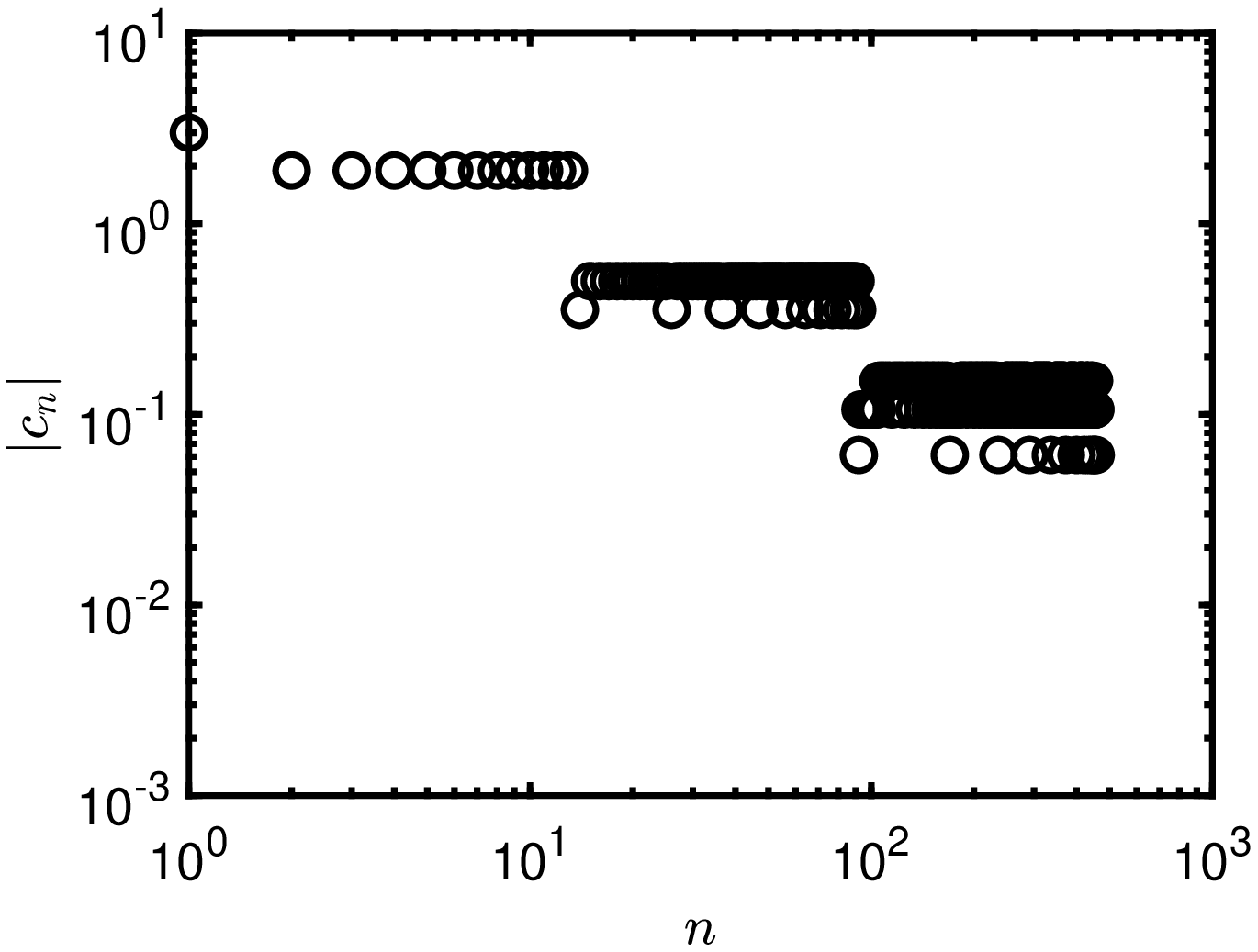}}
     \subfigure[Laguerre $|c_n|$]{
    \includegraphics[width = 0.3\textwidth]{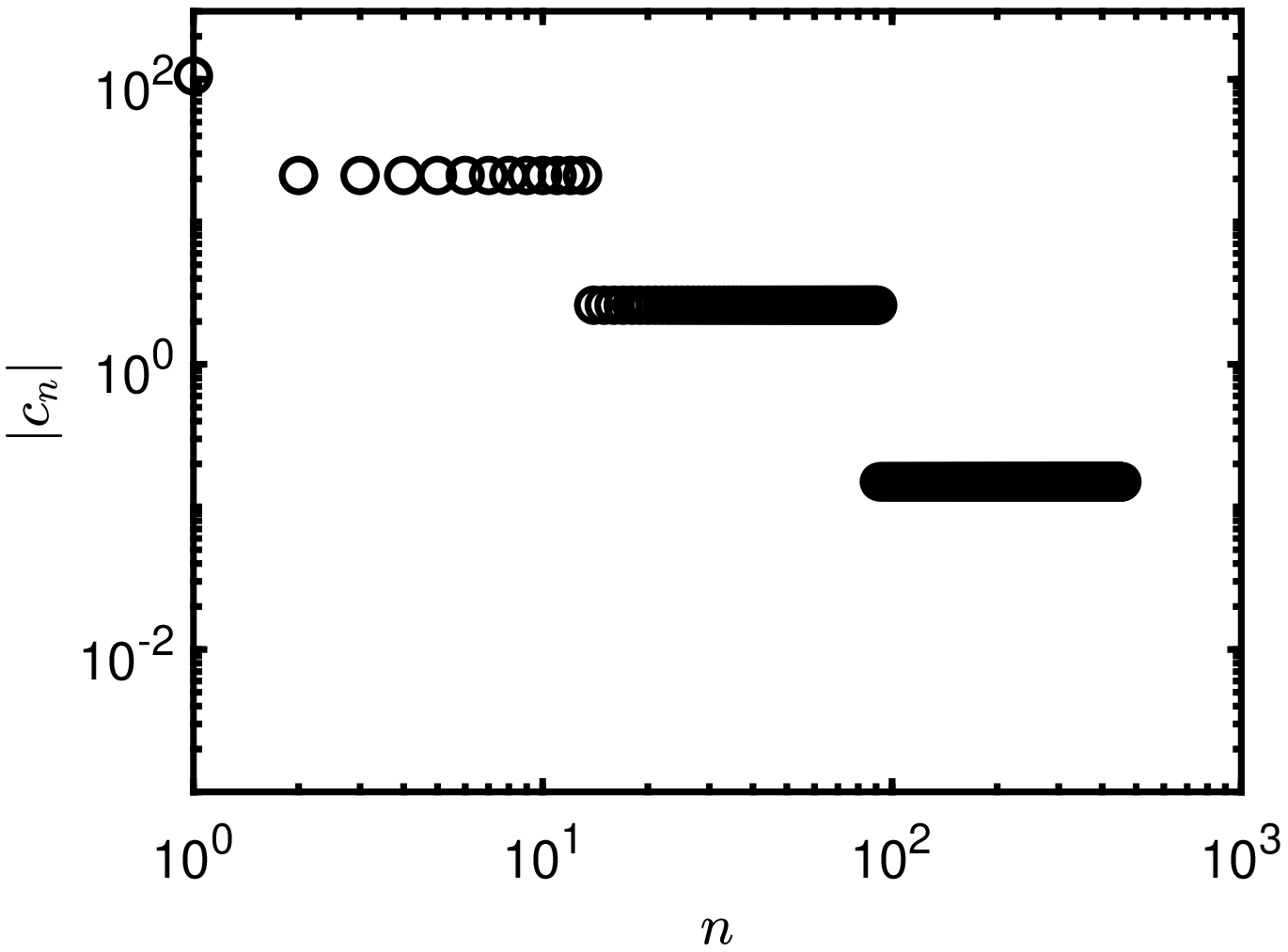}} \\
    \subfigure[Legendre $|\tilde c_n|$]{
    \includegraphics[width = 0.3\textwidth]{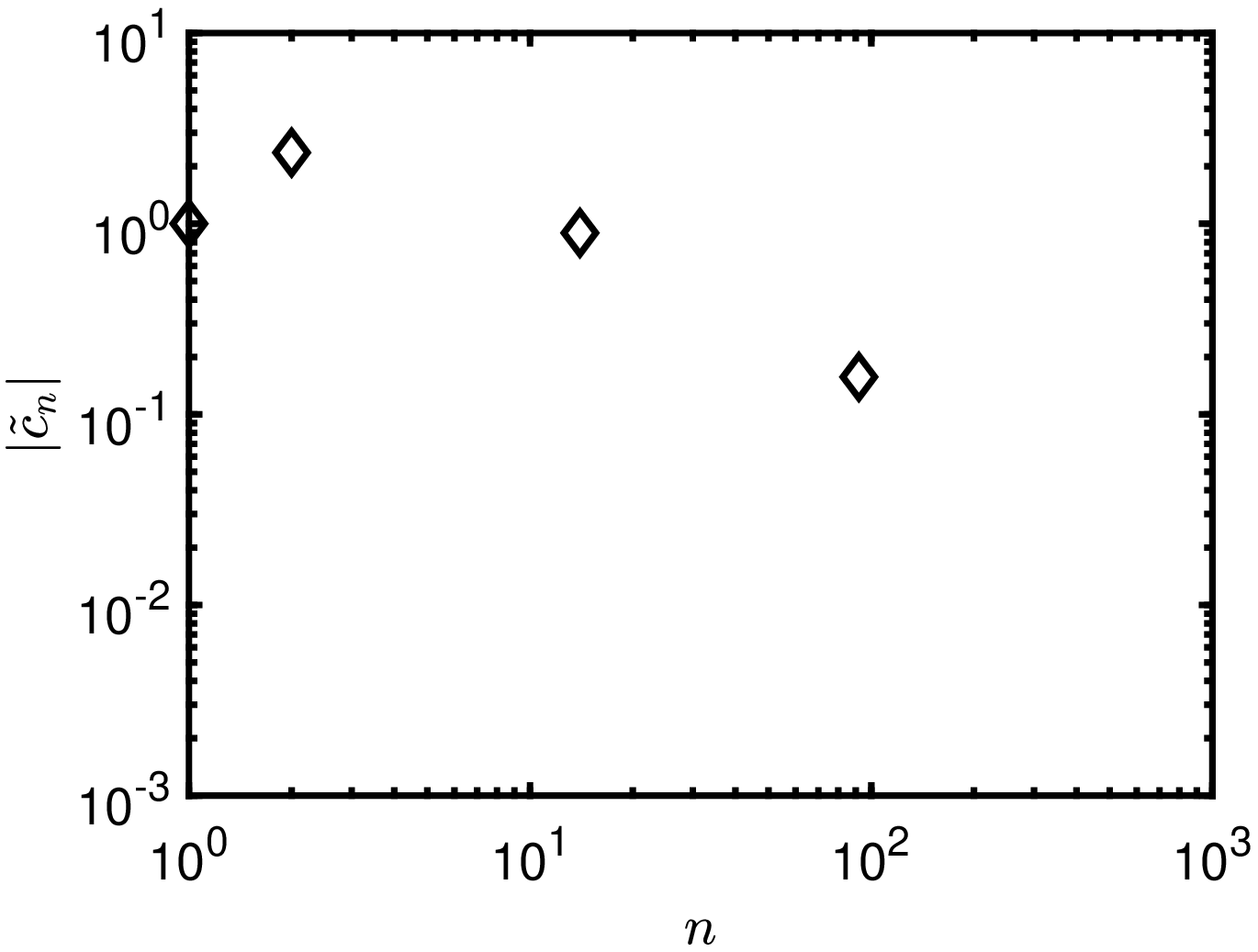}}
    \subfigure[Hermite $|\tilde c_n|$]{
    \includegraphics[width = 0.3\textwidth]{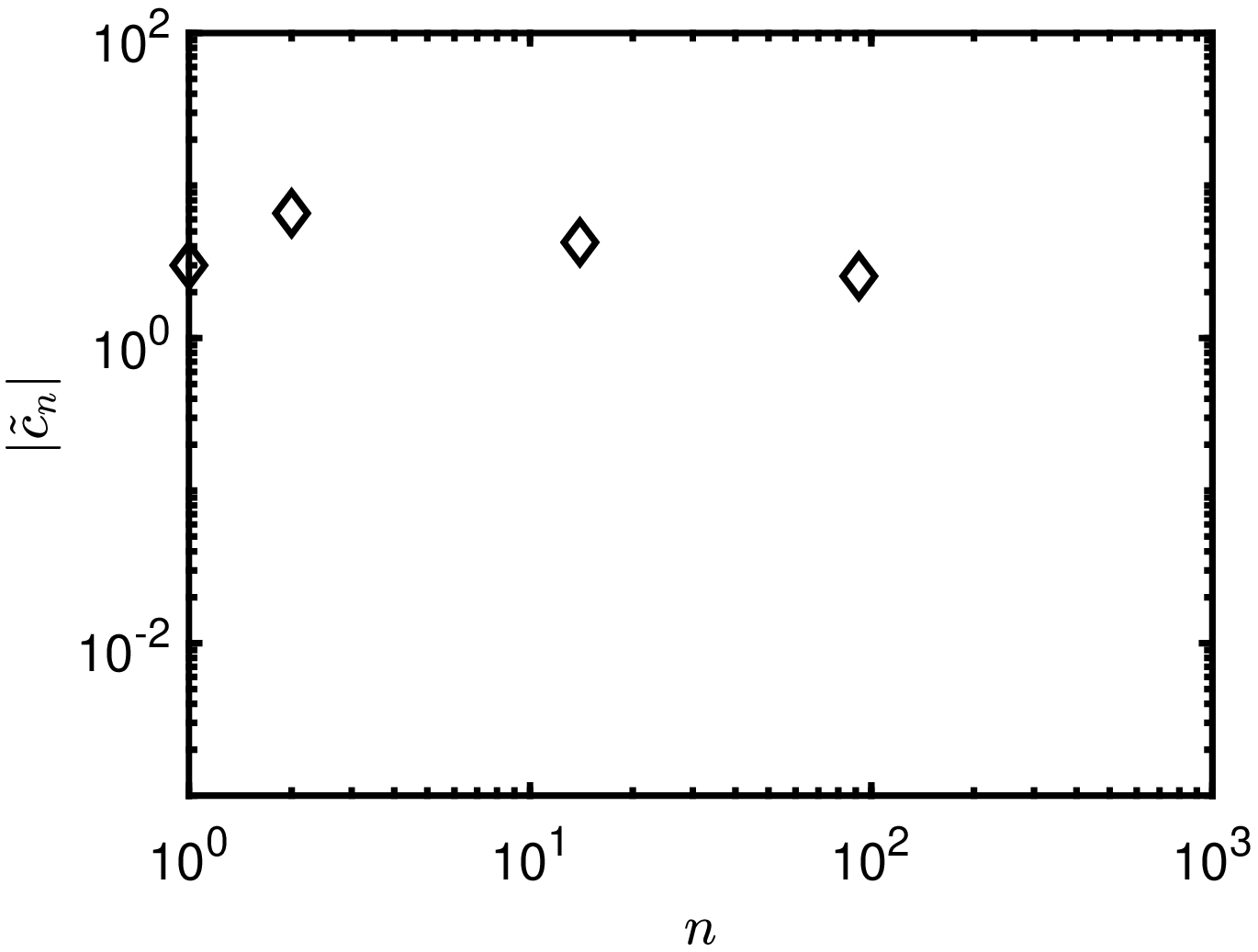}}
     \subfigure[Laguerre $|\tilde c_n|$]{
    \includegraphics[width = 0.3\textwidth]{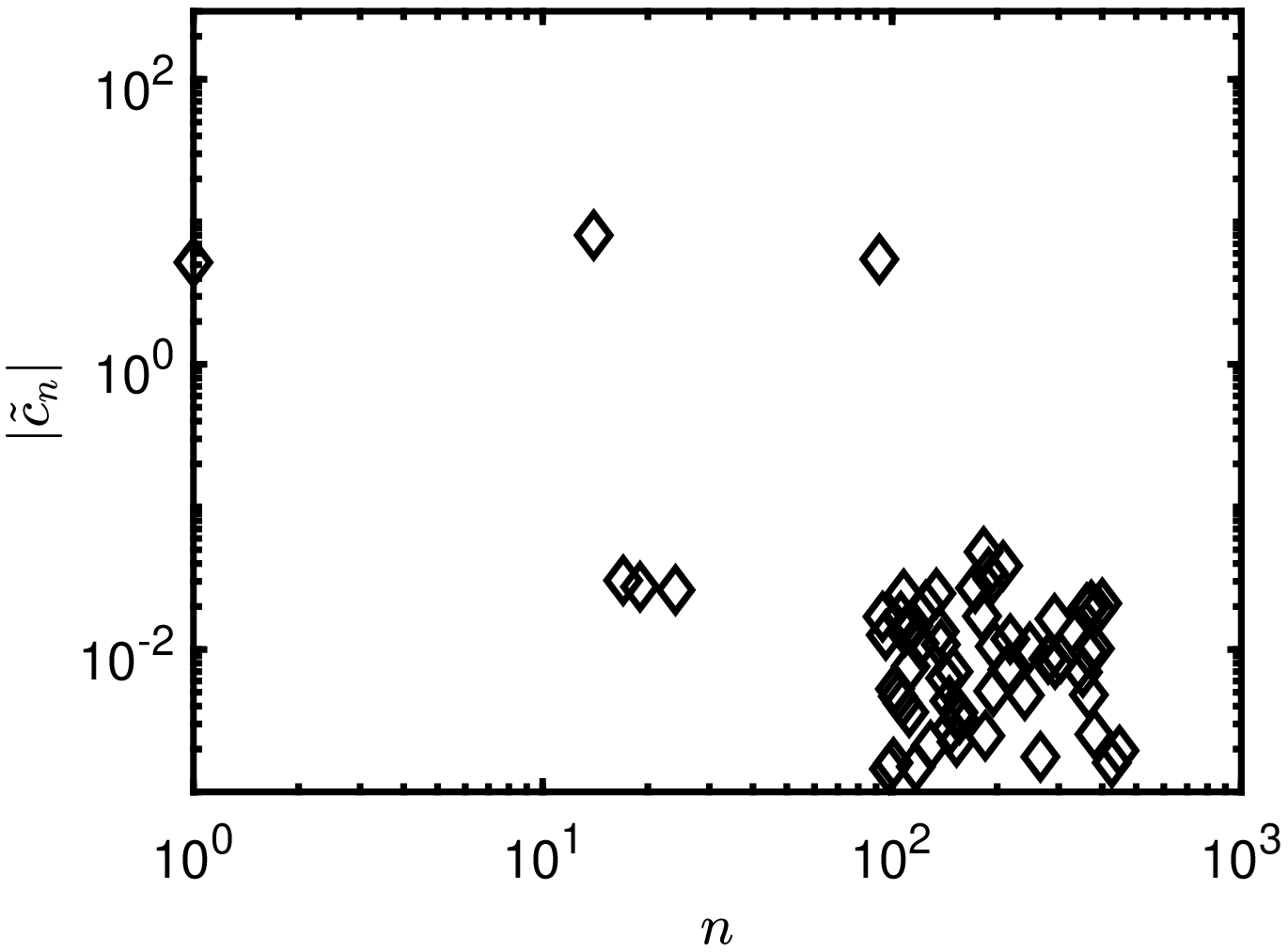}}
    \caption{(Ridge function) Absolute values of exact coefficients $c_{n}$ (firt row) and coefficients
$\tilde{c}_n$ after $9$ rotations with the $\ell_1$-$\ell_2$ minimization  (second row) using $120$ samples. }
    \label{fig:ridge_coef}
\end{figure}

\ref{fig:ridge} also indicates that the standard $\ell_{1}$ minimization without rotation is not effective, as its relative error is close to 50\% even when $M/N$ approaches $0.4$. Our iterative rotation methods with any regularization yield much higher accuracy, especially for a larger $M$ value. 
The reason can be partially explained by \ref{fig:ridge_coef}. Specifically,
the plots (a), (b) and (c) of \ref{fig:ridge_coef} are about the absolute
values of exact coefficients $|c_{n}|$ of Legendre, Hermite, and Laguerre polynomials,
while the plots (d), (e) and (f) show corresponding coefficients
$|\tilde{c}_{n}|$ after 9 iterations with the $\ell_1$-$\ell_2$ minimization
using 120 samples randomly chosen from the $100$ independent experiments; we exclude $\tilde{c}_{n}$ whose absolute value is
smaller than $10^{-3}$, since they are sufficiently small (more than two magnitudes smaller than the dominating
ones). As demonstrated in Figure~\ref{fig:ridge_coef}, the iterative updates on the rotation matrix significantly sparsifies the representation of $u$; and as a result, the efficiency of CS methods is substantially enhanced. 
Moreover, $\tilde c_n$ for the Laguerre polynomial is not as
sparse as the ones for the Legendre and Hermite polynomials. Consequently, the Laguerre polynomial shows less accurate results in
\ref{fig:ridge}(c) compared with other two polynomials in \ref{fig:ridge}(a) and (b). The phenomenon  can be partially explained by the coherence of
$\tensor\Psi$ for different polynomials, i.e., the coherence in the Laguerre case is  much larger
 ($>0.94$ before rotation and $>0.99$ after rotation) than the other
two, thus making  any sparse regression algorithms less effective. For more detailed discussion on coherence, please refer to 
Table~\ref{tab:coherence_Example_ridge} and
 Section~\ref{sect:disc}.

\subsection{Elliptic Equation}\label{sect:elliptic}
Next we consider a one-dimensional elliptic differential equation with a random 
coefficient~\cite{DoostanO11, YangLBL16}:
\begin{equation}\label{eq:ellip}
\begin{aligned}
-\frac{d}{dx} \left( a(x;\bx)\frac{d u(x;\bx)}{dx} \right) = 1, & \quad x \in (0,1) \\
u(0) = u(1) = 0, &
\end{aligned}
\end{equation}
where $a(x;\bx)$ is a log-normal random field based on Karhunen-Lo\`{e}ve (KL) 
expansion:
\begin{equation}\label{eq:kl}
a(x;\bx) = a_0(x) +
\exp\left(\sigma\sum_{i=1}^d\sqrt{\lambda_i}\phi_i(x)\xi_i\right), 
\end{equation}
$\{\xi_i\}$ are i.i.d.~random variables, and
$ \{\lambda_{i}, \phi_{i}(t)\}_{i=1}^{d}$  are eigenvalues/eigenfunctions (in a
descending order in $\lambda_i$) of an exponential covariance kernel:
\begin{equation}\label{eq:exp_kernel}
C(x,x') = \exp\left(\dfrac{|x-x'|}{l_c}\right).
\end{equation}
The value of 
$\lambda_i$ and the analytical expressions for $\phi_i$ are given in~\cite{JardakSK02}. We set 
$a_0(x)\equiv 0.1, \sigma=0.5, l_c=0.2$ and $d=15$ such that
$\sum_{i=1}^d\lambda_i > 0.93\sum_{i=1}^{\infty}\lambda_i$.  For each input 
sample $\bx^q$, the solution of the 
deterministic elliptic equation can be obtained by \cite{YangK13}:
\begin{equation}\label{eq:ellip_sol}
u(x) = u(0) + \int_0^x \dfrac{a(0)u(0)'-y}{a(y)}\dif y.
\end{equation}
By imposing the boundary condition $u(0)=u(1)=0$, we can compute $a(0)u(0)'$ as
\begin{equation}\label{eq:ellip_const}
a(0)u(0)' = \left(\int_0^1 \dfrac{y}{a(y)}\dif y\right) \Big /
            \left(\int_0^1 \dfrac{1}{a(y)}\dif y\right).
\end{equation}
The integrals in Eqs.~\eqref{eq:ellip_sol} and~\eqref{eq:ellip_const} are
obtained by highly accurate numerical integration. For this example, we choose 
the quantity of interest to be $u(x;\bx)$ at $x=0.35$. We aim to build a 
third-order Legendre (or Hermite) polynomial expansion which includes $N=816$
basis functions. The relative error is approximated by a level-$6$ sparse grid 
method. The parameters $(\bar\lambda, \bar\rho)$ are given 
in~\ref{tab:parameter elliptic}.

\begin{table}[t]
    \centering
    \caption{Parameters $(\bar\lambda,\bar\rho)$ for different regularization models in the case of elliptic equation. }
    \label{tab:parameter elliptic}
    \begin{tabular}{c|c c| c c}
    \hline \hline
    UQ setting    & \multicolumn{2}{c|}{ Legendre }    & \multicolumn{2}{c}{ Hermite }      \\ 
        & $\bar\lambda$ & $\bar\rho$    & $\bar\lambda$ & $\bar\rho$\\
    \hline
        $\ell_1$    & $ 6 \times 10^{-4}$   & $1\times 10^{-1}$ & $1.3\times 10^{-3}$   & $3 \times 10^{0}$ \\
        $l_{1/2}$   & $ 8 \times 10^{-5}$   & $1\times 10^{2}$  & $3 \times 10^{-4}$    & $1.2 \times 10^{2}$  \\
        TL$1$       & $1\times 10^{-8}$     & $1\times 10^{-7}$ & $1\times 10^{-8}$     & $1\times 10^{-7}$  \\
        ERF         & $2\times 10^{-2}$     & $2\times 10^{4}$  & $9 \times 10^{-3}$    & $1.3 \times 10^{3}$  \\
        $\ell_1 - \ell_2$ & $1\times 10^{-3}$ & $6\times 10^{-2}$& $1.9\times 10^{-3}$  & $2.3\times 10^{-2}$ \\
    \hline
    \end{tabular}
\end{table}

\begin{figure}[t]
  \centering
  \subfigure[Legendre]{
    \includegraphics[width=0.3\textwidth]{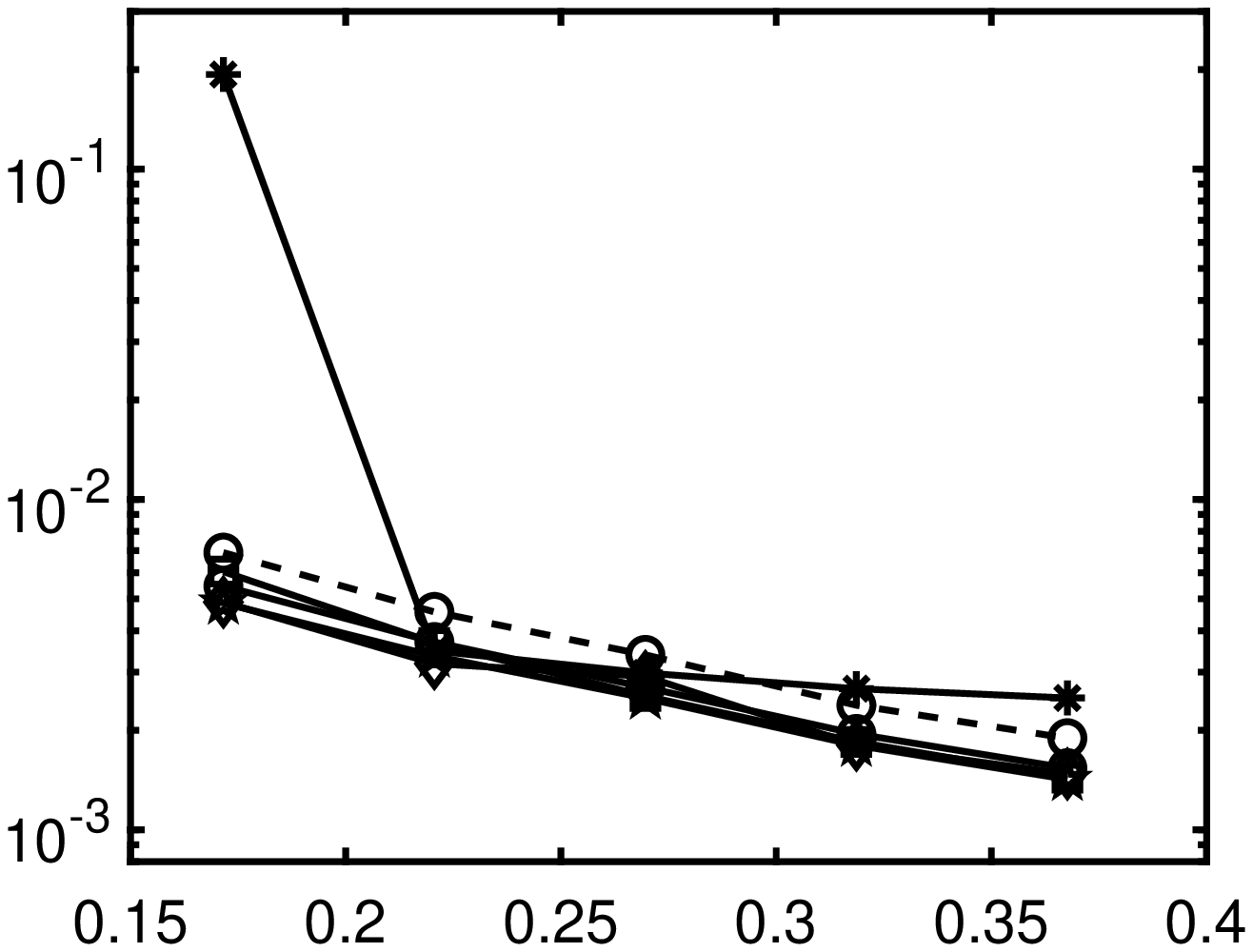}}
  \quad
    \subfigure[Hermite]{
     \includegraphics[width=0.3\textwidth]{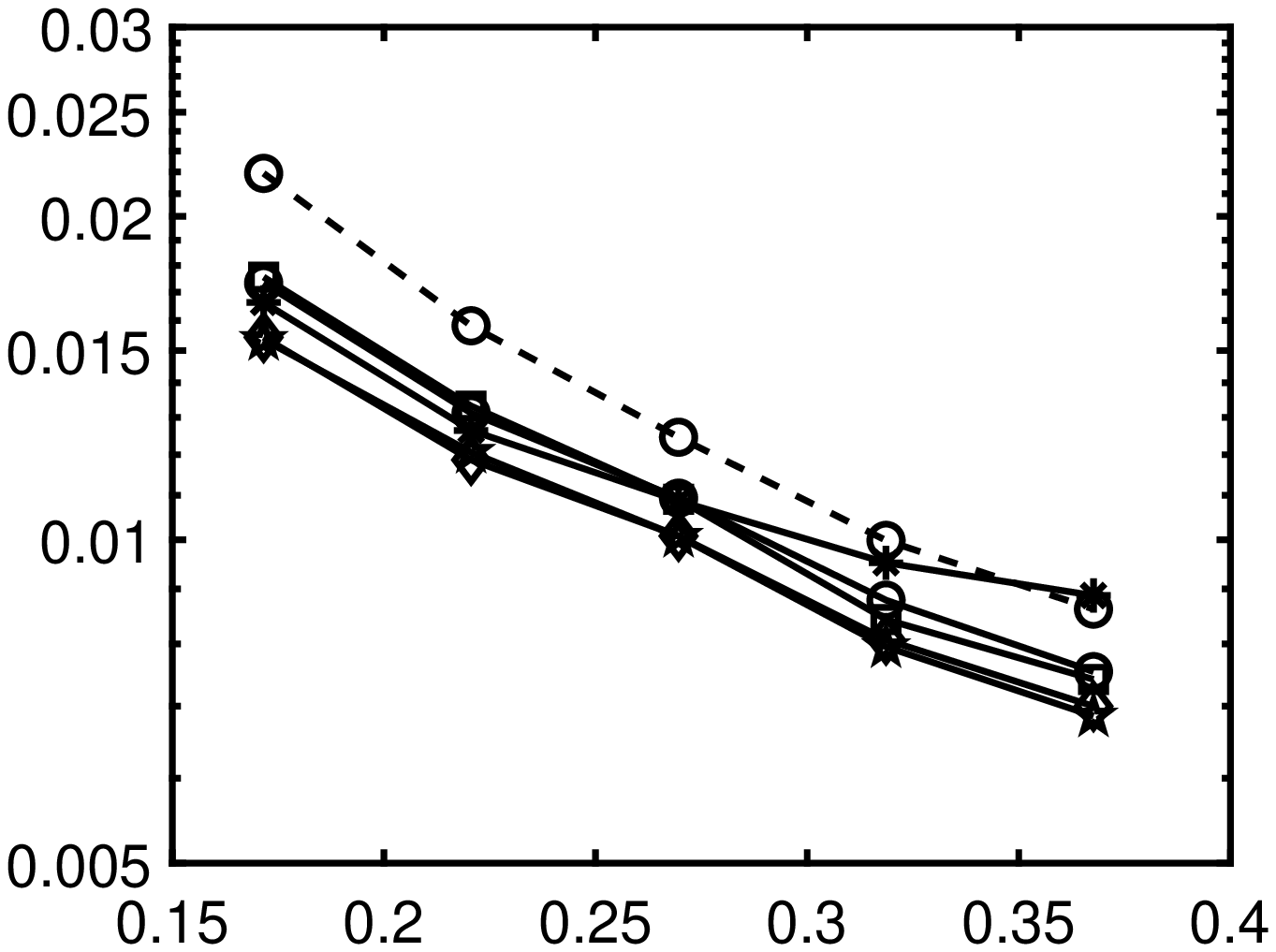}}
    \caption{Relative error of Legendre and Hermite polynomial expansions for an elliptic equation against ratio $M/N$.  Solid lines are for experiments with rotations applied whereas dashed line is a reference of $\ell_{1}$ test result without rotation. `` $\circ$'' is the marker for $\ell_{1}$ minimization, `` $*$'' is the marker for $\ell_{1/2}$, `` $\square$'' is the marker for TL1, `` $\star$'' is the marker for ERF, `` $\Diamond $'' is the marker for $\ell_{1} - \ell_{2}$.}
    \label{fig:elliptic}
\end{figure}

Relative errors of the Legendre and Hermite polynomial expansions are
presented in~\ref{fig:elliptic}. All the methods with rotation perform almost
the same, except that $\ell_{1/2}$ yields unstable results, especially when
sample size is small. In this case, we do not observe significant improvement of the nonconvex regularizations over the convex $\ell_1$ model, as opposed to \ref{fig:ridge}. $\ell_{1}$-$\ell_{2}$ and ERF perform the best for both Legendre and Hermite polynomials.
In this case, the solution does not have an underlying low-dimensional structure
under rotation as in the previous example and the truncation error exists, which is common in practical
problems. This is why the improvement by the rotational method is minor. 

\subsection{Korteweg-de Vries equation}\label{sect:kdv}
\begin{figure}[t]
  \centering
  \subfigure[Legendre]{
    \includegraphics[width = 0.4\textwidth]{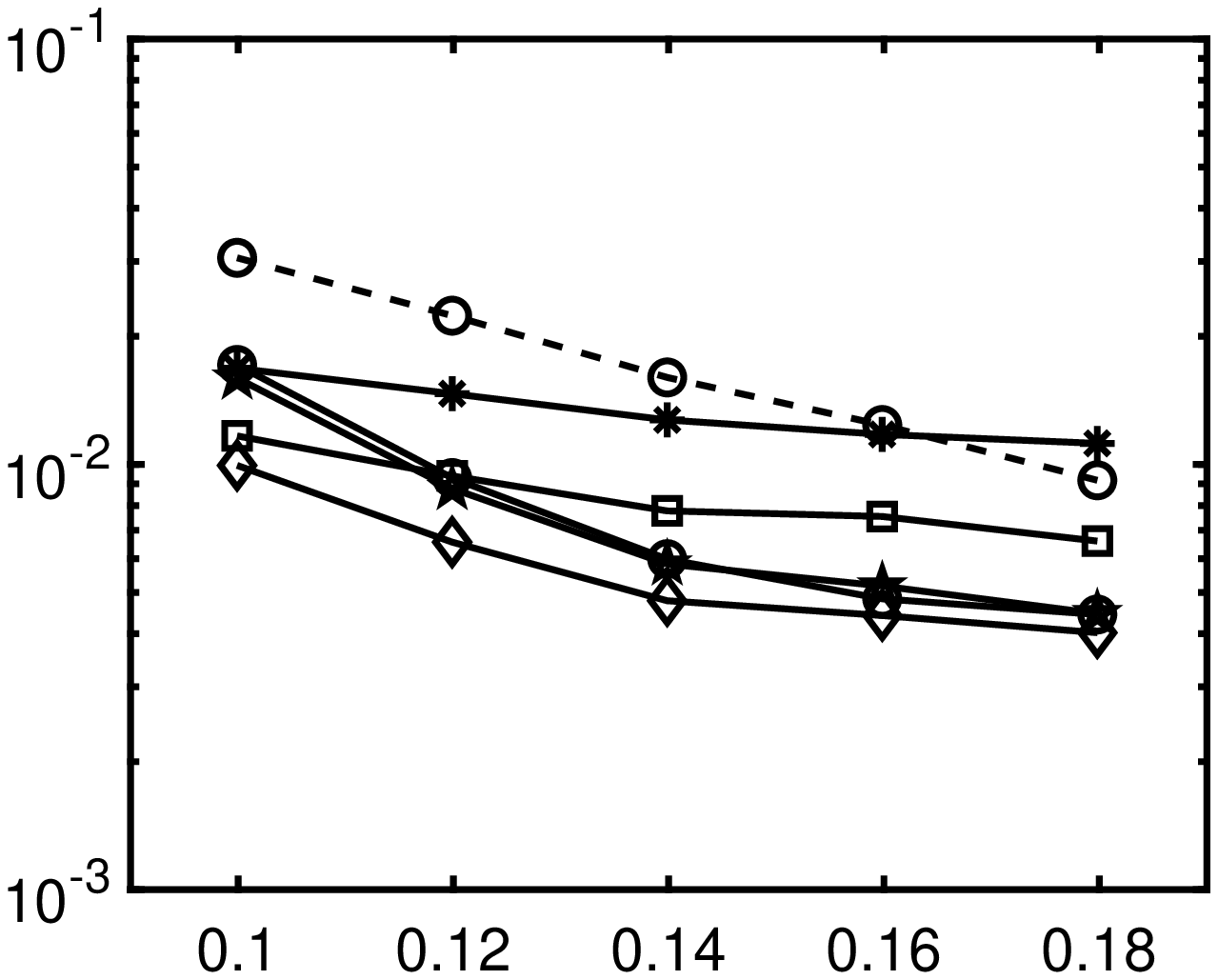}}\qquad
    \subfigure[Hermite]{
    \includegraphics[width = 0.4\textwidth]{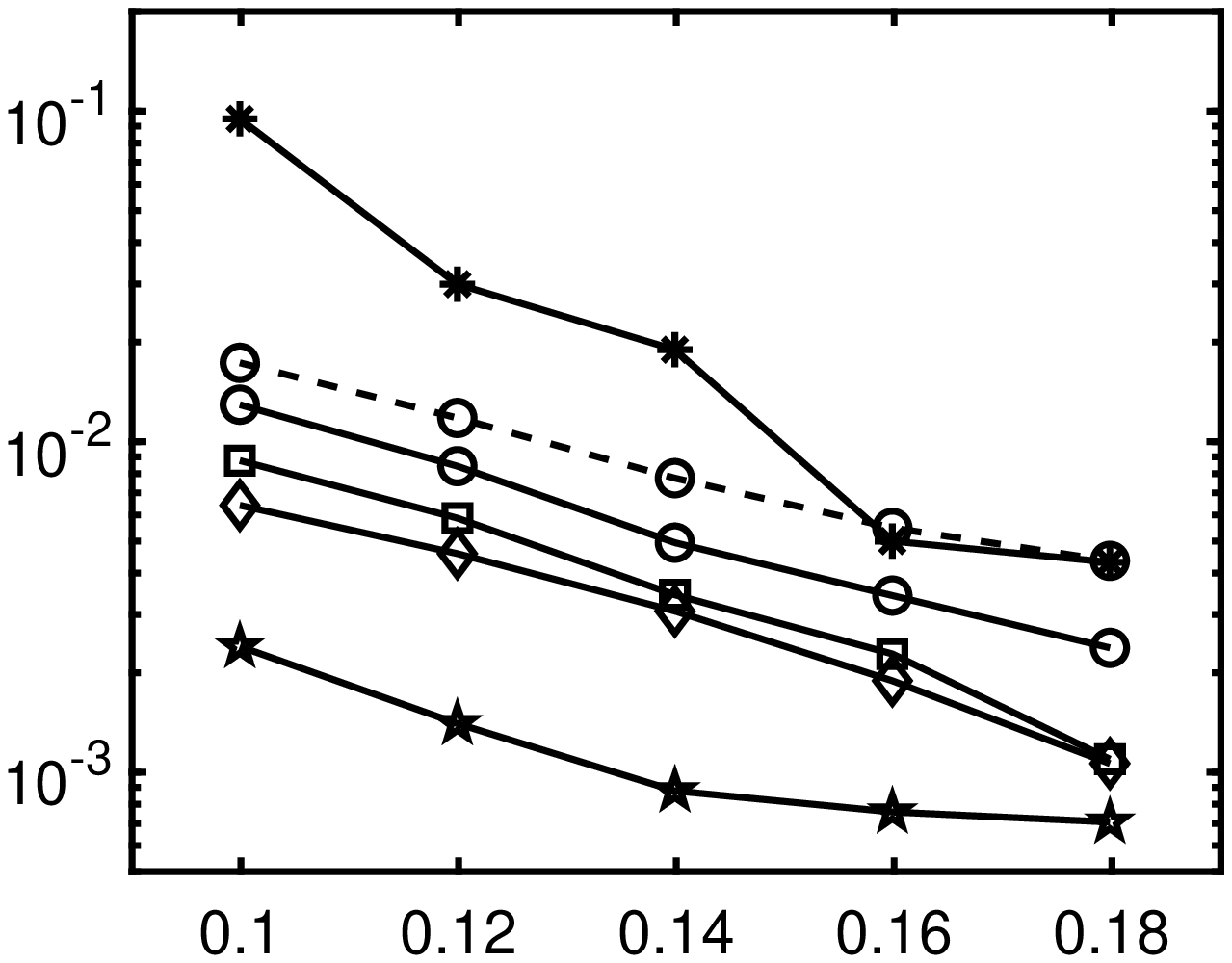}}
    \caption{(KdV equation) Relative error of Legendre and Hermite polynomial expansions for KdV equation against ratio $M/N$. Solid lines are for experiments with rotations applied whereas dashed line is a reference of $\ell_{1}$ test result without rotation. `` $\circ$'' is the marker for $\ell_{1}$ minimization, `` $*$'' is the marker for $\ell_{1/2}$, `` $\square$'' is the marker for TL1, `` $\star$'' is the marker for ERF, `` $\Diamond $'' is the marker for $\ell_{1} - \ell_{2}$.}
    \label{fig:KdV}
\end{figure}

As an example  of  a more complicated and nonlinear differential equation, we consider the Korteweg-de Vries (KdV) equation with time-dependent additive noise, 
\begin{equation}
    \begin{array}{l}
        u_{t}(x, t ; \bm{\xi})-6 u(x, t ; \bm{\xi}) u_{x}(x, t ; \bm{\xi})+u_{x x x}(x, t ; \bm{\xi})=f(t ; \bm{\xi}), \quad x \in(-\infty, \infty) \\
        u(x, 0 ; \bm{\xi})=-2 \operatorname{sech}^{2}(x).
\end{array}
\end{equation}
Here $f(t ; \h{\xi})$ is modeled as a random field represented by the   Karhunen-Lo\`{e}ve   expansion:
\begin{align}
    f(t ; \bm\xi)=\sigma \sum_{i=1}^{d} \sqrt{\lambda_{i}} \phi_{i}(t) \xi_{i},
\end{align}
where $\sigma$ is a constant and $ \{\lambda_{i}, \phi_{i}(t)\}_{i=1}^{d}$  are
eigenvalues/eigenfunctions (in a descending order) of the exponential covariance
kernel Eq.~\eqref{eq:exp_kernel}.
We set $l_c = 0.25$ and $d = 10$ in \eqref{eq:exp_kernel} such that $\sum_{i=1}^{d} \lambda_{i} > 0.96\sum_{i=1}^{\infty} \lambda_{i}$. Under this setting, we have an analytical  solution given by
\begin{align}
    u(x, t ; \bm{\xi})=\sigma \sum_{i=1}^{d} \sqrt{\lambda_{i}} \xi_{i} \int_{0}^{t} \phi_{i}(y) \mathrm{d} y-2 \operatorname{sech}^{2}\left(x-4 t+6 \sigma \sum_{i=1}^{d} \sqrt{\lambda_{i}} \xi_{i} \int_{0}^{t} \int_{0}^{z} \phi_{i}(y) \mathrm{d} y \mathrm{d} z\right).
    \label{eqn:KdV Soliton int}
\end{align}
We choose QoI to be $u(x, t ; \h{\xi})$ at $x = 6, t = 1,$ and $\sigma = 0.4$. Thanks to analytical expressions of $ \phi_{i}(x)$, we can compute the integrals in  \eqref{eqn:KdV Soliton int} with high accuracy.
Denote
\begin{align}
    A_{i}&=\sqrt{\lambda_{i}} \int_{0}^{1} \phi_{i}(y) \mathrm{d} y \quad \mbox{and} \quad B_{i}=\sqrt{\lambda_{i}} \int_{0}^{1} \int_{0}^{z} \phi_{i}(y) \mathrm{d} y \mathrm{d} z, \quad i=1,2, \cdots, d,
\end{align}
 the analytical solution can be written as
\begin{align}
    \left.u(x, t ; \bm\xi)\right|_{x=6, t=1}=\sigma \sum_{i=1}^{d} A_{i} \xi_{i}-2 \operatorname{sech}^{2}\left(2+6 \sigma \sum_{i=1}^{d} B_{i} \xi_{i}\right).
\end{align}
We use a fourth-order gPC expansion to approximate the solution, i.e., $p = 4$, and the number of gPC basis functions is $N = 1001$. The experiment is repeated $50$ times to compute the average relative errors for each gPC expansion. 
Parameters chosen for different regularizations are given in~\ref{tab:parameter kdv}. Relative errors of the Legendre and Hermite polynomial expansions are presented in~\ref{fig:KdV}, which illustrates the combined method of iterative rotation and nonconvex minimization outperforms the simple $\ell_{1}$ approaches. 
The coherence $\mu$ of $\tensor\Psi$ for Legendre polynomial is
around 0.6/0.85 before/after the rotation, while it becomes over 0.92 for the Hermite
polynomial. In such highly coherent regime, 
$\ell_{1/2}$ does not work very well, and other 
nonconvex regularizations, i.e.,   ERF, TL1 and $\ell_1$-$\ell_2$, perform better
than $\ell_1$ in the Hermite polynomial case, especially ERF.

\begin{table}[t]
    \centering
    \caption{Parameters $(\bar\lambda,\bar\rho)$ for different regularization models in the case of KdV function. } 
    \label{tab:parameter kdv}
    \begin{tabular}{c|c c |c c}
    \hline \hline
    Method    & \multicolumn{2}{c|}{ Legendre }    & \multicolumn{2}{c}{ Hermite }      \\ 
        & $\bar\lambda$ & $\bar\rho$    & $\bar\lambda$ & $\bar\rho$\\
    \hline
        $\ell_1$     & $1\times 10^{-4}$ & $1\times 10^{-2}$     & $1\times 10^{-2}$ & $4\times 10^{-4}$ \\
        $l_{1/2}$    & $5 \times 10^{-3}$ & $1.6\times 10^{1}$    & $1.8 \times 10^{-2}$ & $6 \times 10^{-1}$  \\
        Transformed $\ell_1$  & $1\times 10^{-8}$ & $1\times 10^{-7}$ & $1\times 10^{-8}$ & $1\times 10^{-7}$  \\
        ERF  & $1.5\times 10^{-1}$ & $1.6\times 10^{3}$   & $1 \times 10^{0}$ & $1.2 \times 10^{1}$  \\
        $\ell_1 - \ell_2$ & $1\times 10^{-4}$ & $1\times 10^{-2}$ & $1\times 10^{-4}$ & $1\times 10^{-2}$ \\
    \hline
    \end{tabular}
\end{table}

\subsection{High-dimensional function} \label{sect:HD}
\begin{figure}[t]
  \centering
  \subfigure[Legendre]{
    \includegraphics[width = 0.4\textwidth]{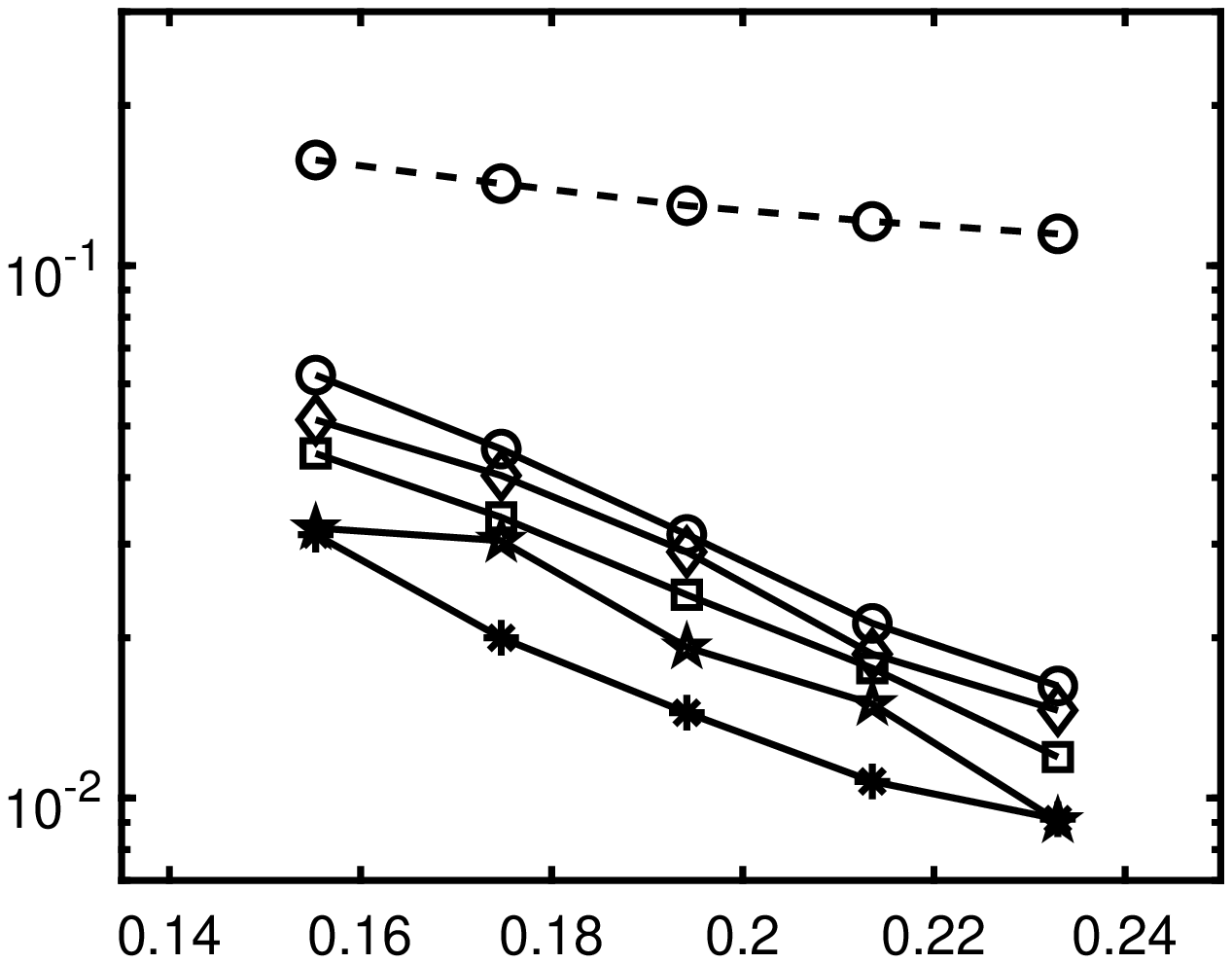}  } \qquad
    \subfigure[Hermite]{
    \includegraphics[width = 0.4\textwidth]{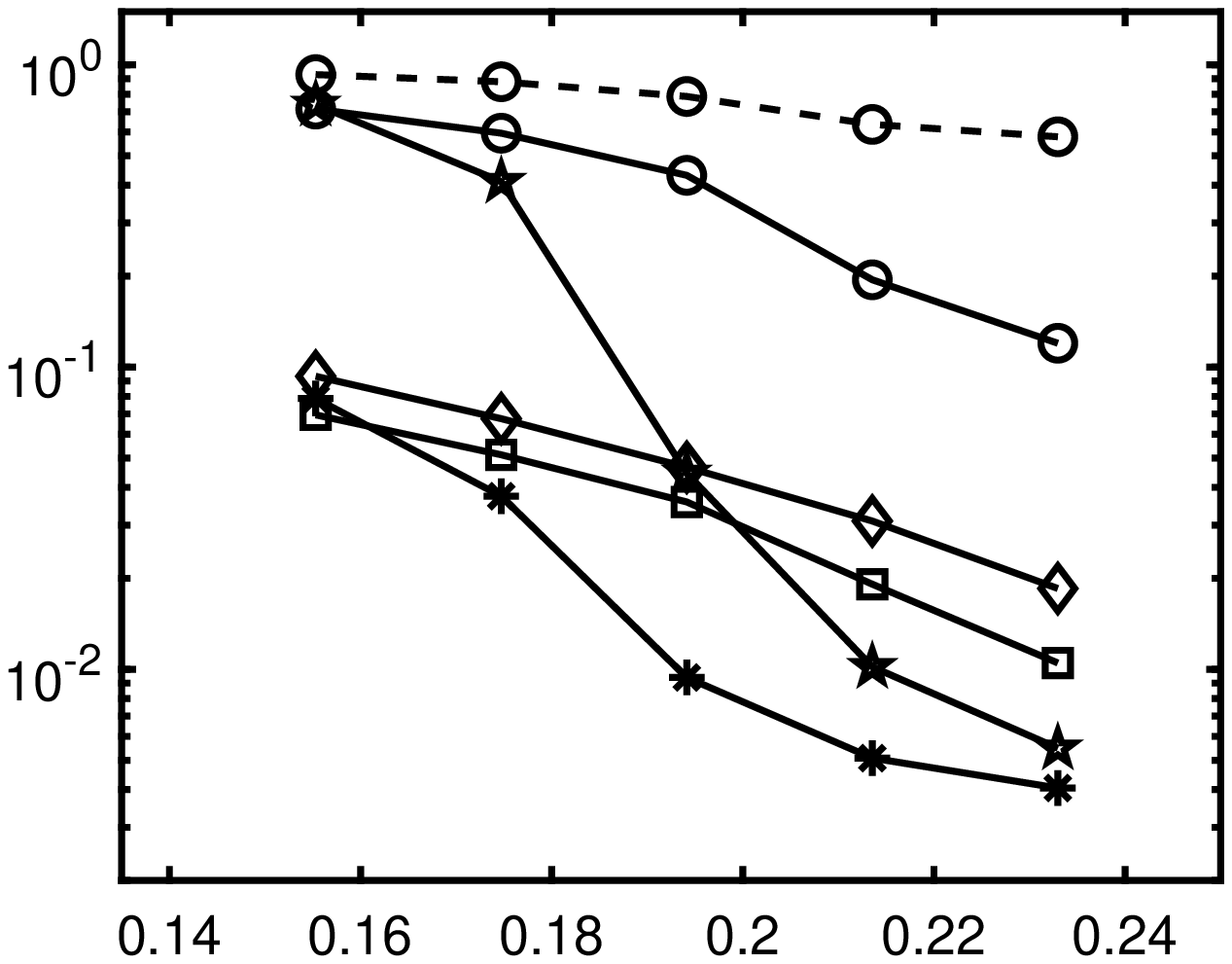}}
    \caption{(High-dimensional function) Relative error of Legendre and Hermite polynomial expansions for the high-dimensional function against sampling ratio $M/N$. Solid lines are for experiments with rotations applied whereas dashed line is a reference of $\ell_{1}$ test result without rotation. `` $\circ$'' is the marker for $\ell_{1}$ minimization, `` $*$'' is the marker for $\ell_{1/2}$, `` $\square$'' is the marker for TL1, `` $\star$'' is the marker for ERF, `` $\Diamond $'' is the marker for $\ell_{1} - \ell_{2}$. All results with rotations are plotted with a threshold of $10^{-3}$. }
    \label{fig:High dim}
\end{figure}

We illustrate the potential capability of the proposed approach for dealing with higher-dimensional problems, referred to as HD function. Specifically, we select a function similar to the one in Section~\ref{sect:ridge} but with a much higher dimension,
\begin{align}
    u(\bm{\xi})=\sum_{i=1}^{d} \xi_{i}+0.25\left(\sum_{i=1}^{d} \xi_{i} / \sqrt{i}\right)^{2}, \quad d=100.
    \label{eqn:high dim eq}
\end{align}
The total number of basis functions for this example is $N = 5151$. 
The experiment is repeated $20$ times to compute the average relative errors for
each polynomial. Parameters for this set of experiments are given in~\ref{tab:parameter HD}. The results are presented in~\ref{fig:High dim}, 
showing that all nonconvex methods with rotation outperforms the $\ell_{1}$ approach except for ERF under the Hermite basis when sample size is small. Different from previous examples (ridge, elliptic, and KdV), $\ell_{1/2}$  achieves the best result,
as the corresponding coherence is relatively small, around 0.2/0.3 before/after rotation for the Legendre polynomial, and $0.3$ for the Hermite polynomial.
In addition, \ref{fig:High dim} (b) suggests that ERF method is stable with respect to
sampling ratios for the Hermite basis. 

\begin{table}[t]
    \label{tab:parameter HD}
    \centering
    \caption{Parameters $(\bar\lambda,\bar\rho)$ for different regularization models in the case of HD function. } 
    \begin{tabular}{c|c c| c c}
    \hline \hline
    UQ setting   & \multicolumn{2}{c|}{ Legendre }    & \multicolumn{2}{c}{ Hermite }      \\ 
        & $\bar\lambda$ & $\bar\rho$    & $\bar\lambda$ & $\bar\rho$\\
    \hline
        $\ell_1$     & $1.2\times 10^{0}$ & $5\times 10^{-2}$     & $1\times 10^{-2}$ & $1\times 10^{-3}$ \\
        $l_{1/2}$    & $1 \times 10^{-1}$ & $1\times 10^{2}$    & $5 \times 10^{-2}$ & $1 \times 10^{2}$  \\
        TL1  & $1\times 10^{-6}$ & $1\times 10^{-7}$ & $4\times 10^{-6}$ & $1\times 10^{-7}$  \\
        ERF  & $1.5\times 10^{-1}$ & $1.6\times 10^{3}$   & $5 \times 10^{-2}$ & $1 \times 10^{2}$  \\
        $\ell_1 - \ell_2$ & $5\times 10^{-2}$ & $1\times 10^{2}$ & $1\times 10^{-3}$ & $9\times 10^{-2}$ \\
    \hline
    \end{tabular}
\end{table}

\subsection{Discussion}\label{sect:disc}
We intend to discuss the effects of coherence and the number of rotations on
the performance of the $\ell_1$ and other nonconvex approaches. As reported in
\cite{guo2020novel,louYHX14,yinLHX14}, $\ell_1$-$\ell_2$ and ERF methods perform
particularly well for coherent matrices (i.e., large $\mu$) and $\ell_p$
performs well for incoherenct matrices (i.e., small $\mu$), which motivates us
to compute the coherence values and report
in~\ref{tab:coherence_Example_ridge},~\ref{tab:coherence_Example_kdv} using the $\ell_{1}$-$\ell_{2}$ method for ridge and elliptic/KdV/HD, respectively.
Here, we use $\mu$ of each iteration by $\ell_1$-$\ell_2$ as an examples, and its value for other
regularizations (including $\ell_1$) are similar.
Both tables confirm that applying rotation increases the coherence level of the sensing matrix $\tensor \Psi$ except for the Hermite basis. 
As we show in numerical examples,
when the coherence is large (e.g., around 0.9 or even larger in Hermite
polynomial for KdV) $\ell_1$-$\ell_2$
and ERF perform better than the convex $\ell_1$ method.
When the coherence is small (e.g., $\lesssim 0.3$ in the high-dimensional case) 
$\ell_{1/2}$ gives the best results among all the competing methods. 
One the other hand, $\ell_{1/2}$ may lead to unstable and unsatisfactory
results, sometimes even worse than the convex $\ell_1$ method, when the
coherence of the sensing matrix is large. In the extreme case when the coherence
is close to $1$ (e.g., Laguerre in ridge function), the best result is only slightly better than $\ell_{1}$, which seems difficult for any sparse recovery algorithms to succeed.
This series of observations coincide with the empirical performance in CS,
i.e., $\ell_{1/2}$ works the best for incoherent matrices, while
$\ell_1$-$\ell_2$ and ERF work better for coherent cases.



\begin{table}[t]
    \centering
    \caption{Average coherence of matrix $\tensor\Psi$ (size $160\times 455$) for ridge function. } 
    \begin{tabular}{c|c c c}
    \hline\hline
    & & Ridge function  &  \\
    Rotations    &Legendre   &Hermite    &Laguerre   \\ 
    \hline
    0  &0.4692 & 0.7622 & 0.9448 \\
    3  &0.6833 & 0.7527 & 0.9910 \\
    6  &0.6822 & 0.7519 & 0.9911 \\
    9  &0.6762 & 0.7657 & 0.9911 \\
    \hline
    \end{tabular}
    \label{tab:coherence_Example_ridge}
\end{table}

\begin{table}[t]
    \begin{center}
    \caption{Average coherence of matrix $\tensor\Psi$ for ellip
      equation (matrix size $160\times 816$), KdV  (matrix size $160\times 1001$) and HD function (matrix size $1000\times 5151$). } 
    \label{tab:coherence_Example_kdv}
    \begin{tabular}{c|cc|cc|cc}
    \hline\hline
     & \multicolumn{2}{c|}{Elliptic equation}& \multicolumn{2}{c|}{KdV equation}  &\multicolumn{2}{c}{HD function}   \\
    Rotations    &Legendre &Hermite&Legendre &Hermite&Legendre &Hermite \\ 
    \hline
    0  & 0.5014 & 0.7770 & 0.6079    & 0.9121 & 0.2117 & 0.2852\\
    1  & 0.6958 & 0.7830 & 0.8825    & 0.9223 & 0.2580 & 0.3075\\
    2  & 0.6943 & 0.7719 & 0.8487    & 0.9167 & 0.2664 & 0.2956\\
    3  & 0.6896 & 0.7710 & 0.8677    & 0.9214 & 0.2522 & 0.3003\\
    \hline
    \end{tabular}
    \end{center}
\end{table}

\begin{figure}[ht]
  \centering
  \subfigure[Ridge Legendre]{
    \includegraphics[width = 0.4\textwidth]{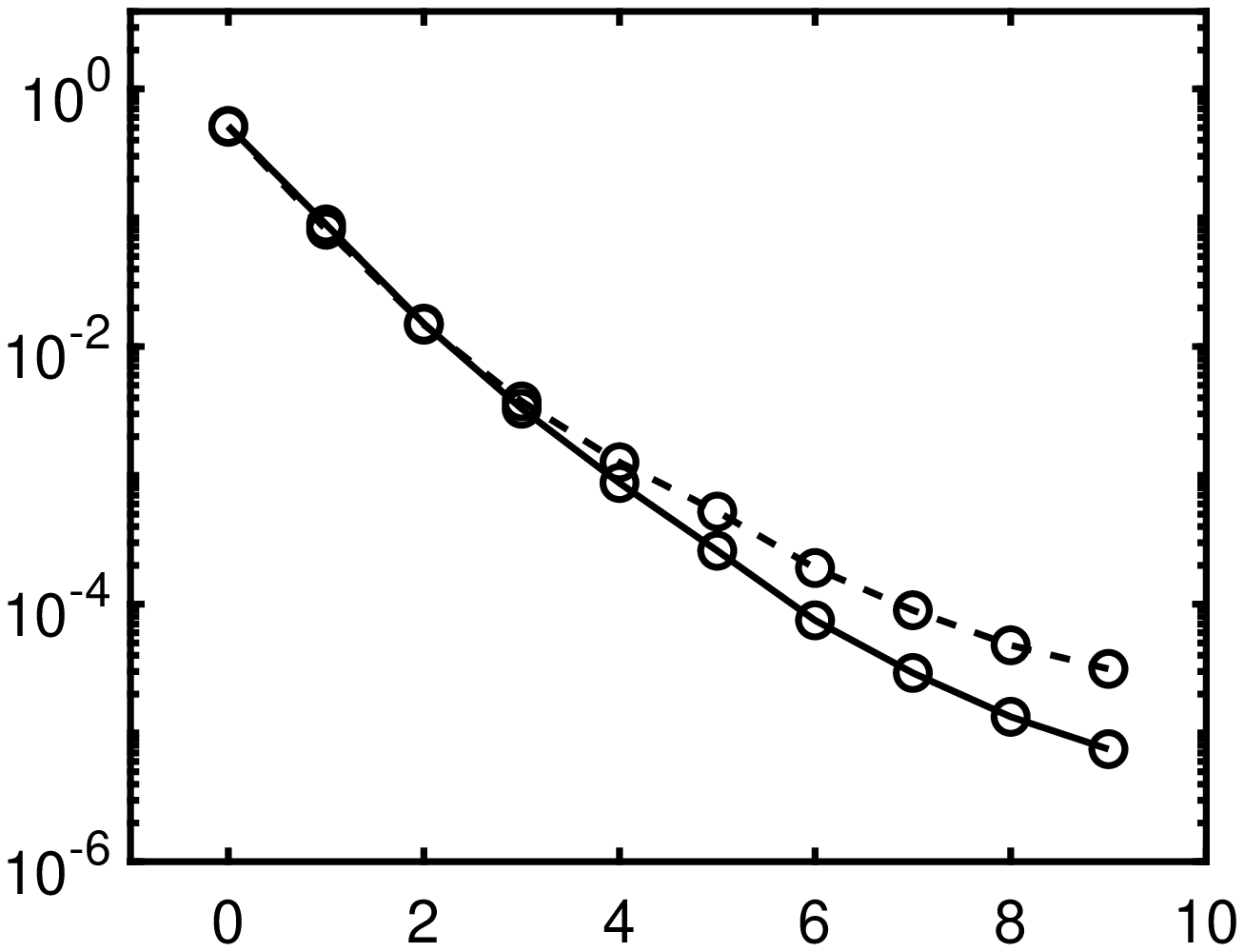} }\qquad
    \subfigure[Ridge Hermite]{
    \includegraphics[width = 0.4\textwidth]{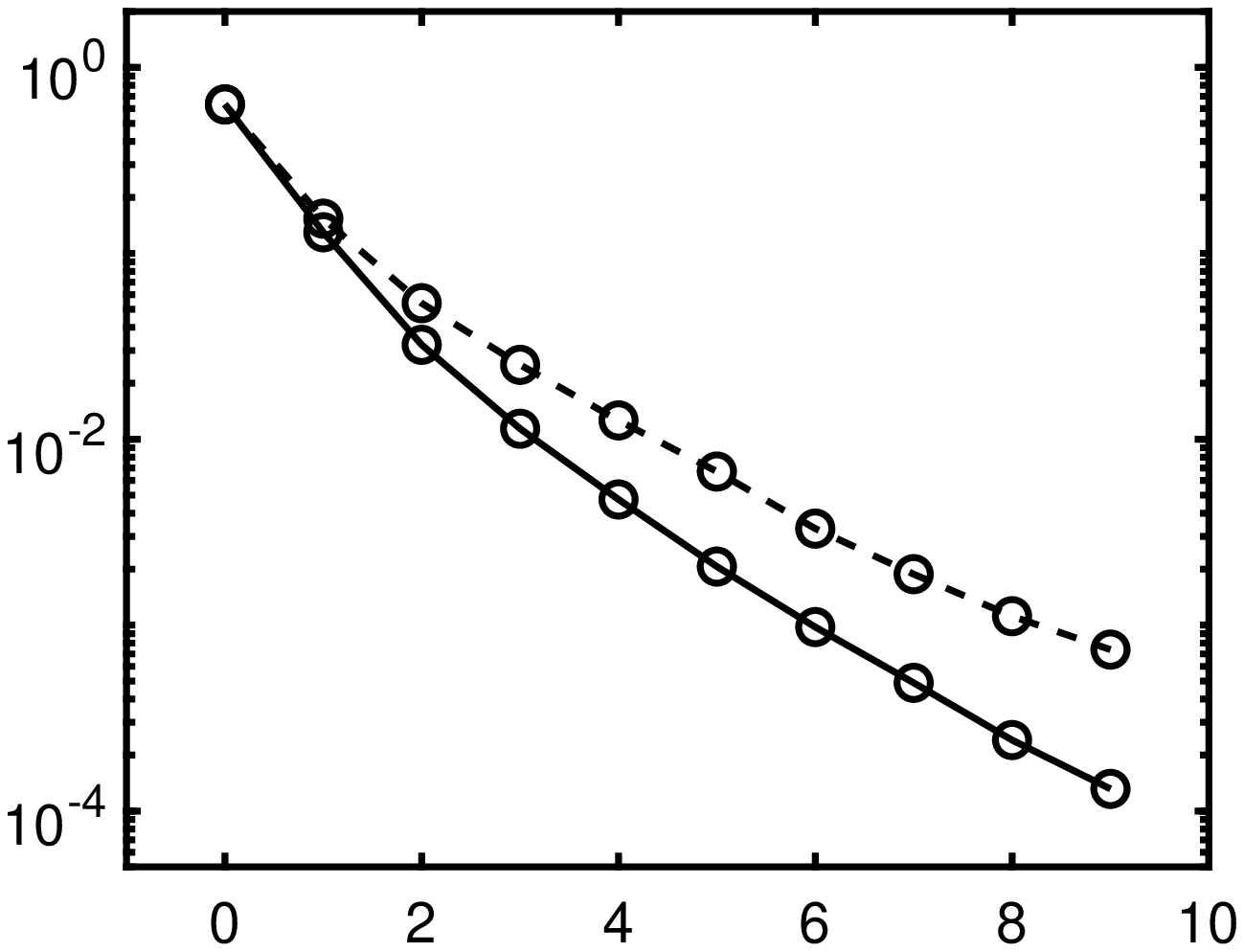}}\\
    \subfigure[KdV Legendre]{
    \includegraphics[width = 0.4\textwidth]{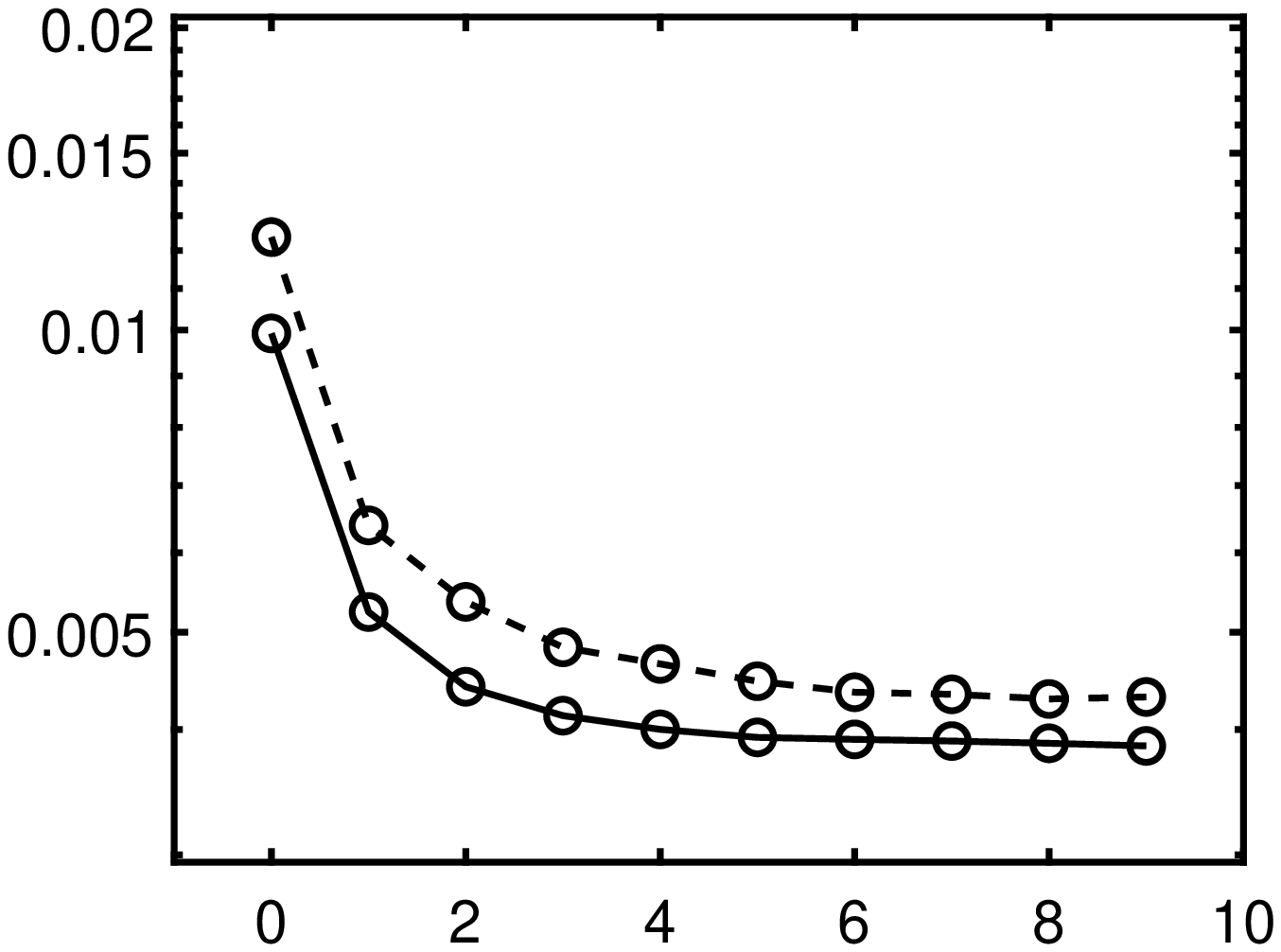}} \qquad
    \subfigure[KdV Hermite]{
    \includegraphics[width = 0.4\textwidth]{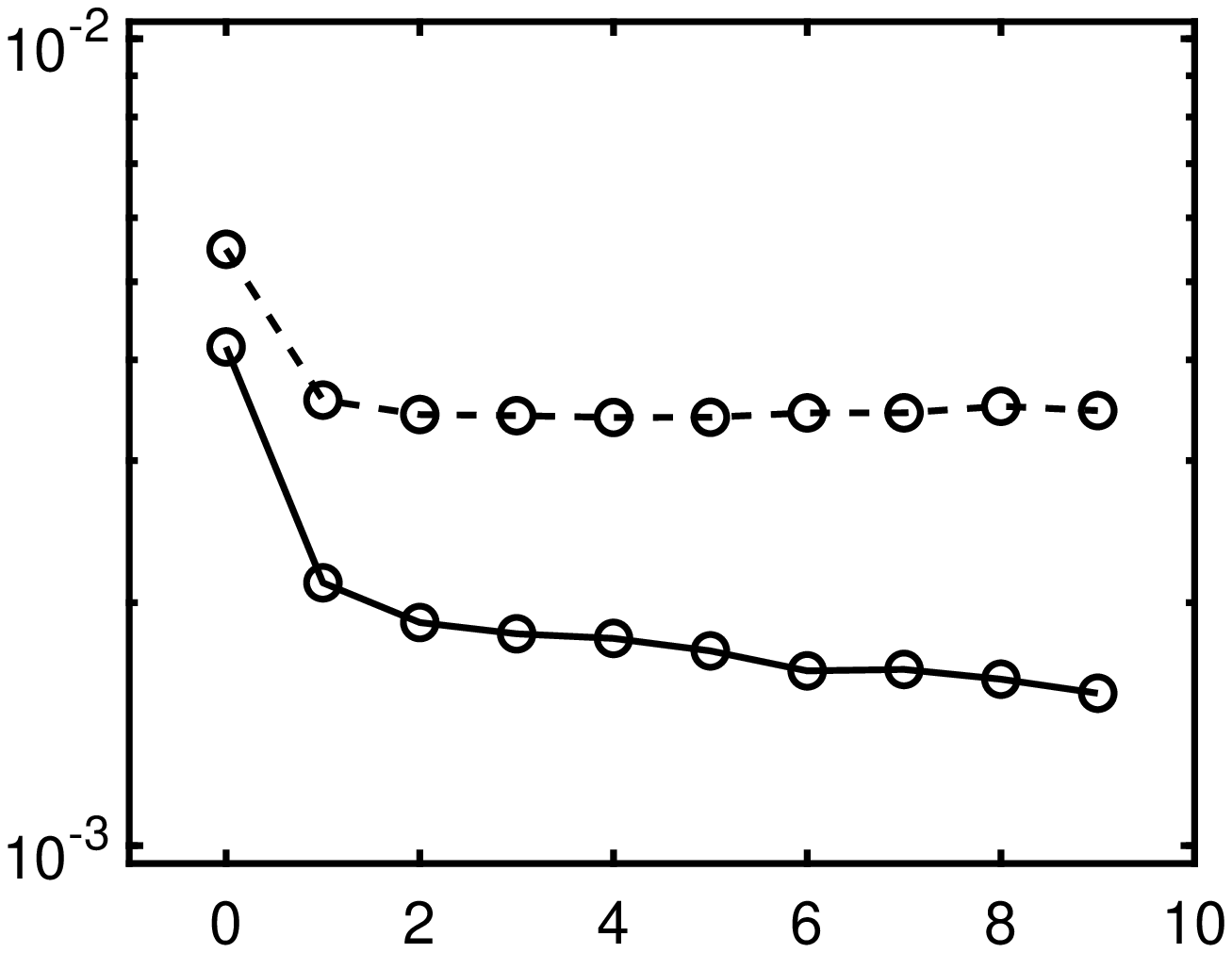}}
  \caption{Relative error vs rotation. The size of $\tensor\Psi$ is $160\times 455$ in the Ridge problem and $160\times 1001$ in the KdV problem. }
    \label{fig:REvsRot}
\end{figure}

We then examine the effect of the number of rotations in~\ref{fig:REvsRot}.
Here we use ridge function and KdV equation as an examples, and show the
results by $\ell_1$-$\ell_2$. For ridge function, more rotations lead to a better accuracy, while it stagnates at 3-5 rotations for the KdV equation. 
This is because the ridge function has very good low-dimensional structure,
i.e., the dimension can be reduced to one using a linear transformation
$\bm\eta=\tensor A\bm\xi$, while the KdV equation does not have this property.
Also, there is no truncation error $\varepsilon(\bm\xi)$ when using the gPC
expansion to represent the ridge function as we use a third order expansion,
while $\varepsilon(\bm\xi)$ exists for the KdV equation. In most practical
problems, the truncation error exists and the linear transform may not yield the
optimal low-dimensional structure to have sparse coefficients of the gPC expansion. Therefore, we empirically set a maximum number of rotations $l_{\max}$ to terminate iterations in the algorithm. 

Finally, we present the computation time in~\ref{tab:time}.  All the experiments are performed on an AMD Ryzen 5 3600, 16 GB RAM machine on Windows 10 1904 and 2004 with MATLAB 2018b.
The major computation comes from two components: one is the  $\ell_{1}$-$\ell_{2}$ minimization and the other is to find the rotation matrix $A$.
The computation complexity for every iteration of the $\ell_{1}$-$\ell_{2}$ algorithm is $O(M^3+M^{2}N),$ which reduces to $O(M^2N)$ as we assume $M\ll N.$  In practice, we choose the maximum outer/inner numbers in Algorithm~\ref{algo:cs_l1l2} as $n_{max} = 10, k_{max} = 2N$ respectively, and hence  the  complexity for $\ell_{1}$-$\ell_{2}$ algorithm is  $O(M^{2}N^{2})$. To find the rotation matrix $A,$ one has to construct a matrix $W$ using \eqref{eq:grad_iter}  with complexity of $O(M^{3}N),$ followed by SVD with complexity of  $O(M^{3}N+M^{2}N^{2}).$ Therefore, the total complexity of our approach is $O(M^{2}N^{2})$ per rotation. We divide the time of Legendre polynomial reported in~\ref{tab:time} by $l_{\max} (MN)^2,$ getting  $1.39e^{-10}, 1.95e^{-10}$, and $0.25e^{-10}.$ As the ratios are of the same order, the empirical results are  consistent with the complexity analysis. 

\begin{table}
    \centering
    \caption{Computation time  per each random realization, averaged over 20 trials. (N/A means a certain case is not available.) } 
    \label{tab:time}
    \begin{tabular}{c c c c c c}
    \hline\hline
    Time (\it{sec.})  & dimension &rotations  &Legendre   &Hermite &Laguerre  \\ 
    \hline
    Ridge function &$160\times455$     &9  &6.53  & 4.31     & 16.19 \\
    Elliptic equation   &$220\times 816$    &3  &5.26 & 11.96   & N/A   \\
    KdV equation   &$160\times 1001$    &3  &15.03   & 14.33   & N/A   \\
    HD function&$1000\times 5151$    &3  &2041.82  & 2102.04  & N/A \\
    \hline
    \end{tabular}
\end{table}

\section{Conclusions} 
\label{sec:conclusion}
In this work, we proposed an alternating direction method to identify a rotation matrix iteratively in order to enhance the sparsity of  gPC expansion, followed by several nonconvex minimization scheme to efficiently identify the sparse coefficients. 
We used a general framework to incorporate any regularization whose proximal
operator can be found efficiently (including $\ell_1$) into the rotational
method.
As such, it improves the accuracy of the compressive sensing method to construct the gPC expansions from a small amount of data.
In particular, the rotation is determined by seeking the directions of maximum
variation for the QoI through SVD of the gradients at different points in the
parameter space. The linear system after rotations becomes ill-conditioned,
specifically more coherent, which motivated us to choose the nonconvex method
instead of the convex $\ell_{1}$ approach for sparse recovery. We conducted
extensive simulations under various scenarios, including a ridge function, an
elliptic equation, KdV equation, and a HD function with Legendre, Hermite, and
Laguerre polynomials, all of which are widely used in practice. Our
experimental results demonstrated that the proposed combination of rotation
estimation and nonconvex methods significantly outperforms the standard $\ell_1$
minimization (without rotation). In different coherence scenarios, there are
different nonconvex regularizations (combined with rotations) that outperforms the rotational CS with the $\ell_1$
approach. Specifically, $\ell_1$-$\ell_2$ and ERF work well for coherent
systems, while $\ell_{1/2}$ excels in incoherent ones, which are
aligned with the observations in CS studies.

\section*{Acknowledgement}
This work was partially supported by NSF CAREER 1846690.
Xiu Yang was supported by the U.S. Department of Energy (DOE), Office of Science, Office of Advanced Scientific Computing Research (ASCR) as part of Multifaceted Mathematics for Rare, Extreme Events in Complex Energy and Environment Systems (MACSER).

\bibliographystyle{siam}
\bibliography{reference}
\end{document}